\documentclass[12pt]{iopart}
\usepackage{color}
\usepackage{graphics}
\usepackage{epsfig}
\usepackage{amssymb}
\usepackage{hyperref}

\begin{document}

\title[]{The TRAPSENSOR Facility: an Open-Ring 7-Tesla Penning Trap for Laser-Based Precision Experiments}

\author{Manuel J~Guti\'errez$^1$, Joaqu\'in~Berrocal$^1$, Juan Manuel~Cornejo$^1$\footnote{Present address: Institut f\"ur Quantenoptik, Leibniz Universit\"at Hannover, Welfengarten 1, 30167 Hannover, Germany}, Francisco~Dom\'inguez$^1$, Jes\'us J~Del Pozo$^1$, I\~nigo~Arrazola$^2$, Javier~Ba\~nuelos$^1$, Pablo~Escobedo$^1$\footnote{Departamento de Electr\'onica y Tecnolog\'ia de Computadores, Universidad de Granada, 18071, Granada, Spain}, Oliver~Kaleja$^{3,4}$, Lucas~Lamata$^2$, Ra\'ul A~Rica$^{1,5}$\footnote{Departamento de F\'isica Aplicada, Universidad de Granada, 18071, Granada, Spain}, Stefan~Schmidt$^1$\footnote{Institut f\"ur Physik, Johannes Gutenberg-Universit\"at Mainz, 55099 Mainz, Germany}, Michael~Block$^{3,4,6}$, Enrique~Solano$^{2,7,8}$ and Daniel~Rodr\'iguez$^{1,5}$\footnote{Corresponding 
author: danielrodriguez@ugr.es} }

\address{$^1$Departamento de F\'isica At\'omica, Molecular y Nuclear, Universidad de Granada, 18071, Granada, Spain}
\address{$^2$Department of Physical Chemistry, University of the Basque Country UPV/EHU, Apartado 644, 48080, Bilbao, Spain}
\address{$^3$Institut f\"ur Kernchemie, Johannes Gutenberg-Universit\"at Mainz, 55099 Mainz, Germany}
\address{$^4$GSI Helmholtzzentrum f\"ur Schwerionenforschung GmbH, 64291, Darmstadt, Germany}
\address{$^5$Centro de Investigaci\'on en Tecnolog\'ias de la Informaci\'on y las Comunicaciones,  Universidad de Granada, 18071, Granada, Spain}
\address{$^6$Helmholtz-Institut Mainz, 55099,  Mainz, Germany}
\address{$^7$IKERBASQUE, Basque Foundation for Science, Maria Diaz de Haro 3, 48013, Bilbao, Spain}
\address{$^8$Department of Physics, Shanghai University, 200444 Shanghai, China}

\vspace{10pt}

\begin{abstract}
The Penning-trap electronic-detection technique that offers the precision and sensitivity requested in mass spectrometry for fundamental studies in nuclear and particle physics has not been proven yet to be universal. This has motivated the construction of a Penning-trap facility aiming at the implementation of a novel detection method, consisting in measuring motional frequencies of singly-charged trapped ions in strong magnetic fields, through the fluorescence photons from the 4s$^2$S$_{1/2}\rightarrow $4p$^2$P$_{1/2}$ atomic transition in $^{40}$Ca$^+$. The key element of this facility is an open-ring Penning trap, built and fully characterized, which is coupled upstream to a preparation Penning trap similar to those built at Radioactive Ion Beam facilities. Motional frequency measurements of trapped ions stored in the open-ring trap have been carried out by applying external dipolar and quadrupolar fields in resonance with the ions' eigenmotions, in combination with time-of-flight identification. The infrastructure to observe the fluorescence photons from $^{40}$Ca$^+$, comprising the twelve laser beams needed in 7~Tesla, and a two-meters long system to register the image in a high-sensitive CCD sensor, has also been successfully tested by observing optically the trapped $^{40}$Ca$^+$ ions. This demonstrates the capabilities of this facility for the proposed laser-based mass-spectrometry technique, and introduces it as a unique platform to perform laser-spectroscopy experiments with implications in different fields of physics.

\end{abstract}


\section{Introduction}

Mass measurements with the highest precision, on stable and exotic nuclei, are performed using Penning traps, outstanding and versatile tools for studying fundamental properties of atoms and ions \cite{Brow1986,Blau2013}. The required relative mass uncertainty $\delta m/m$ differs by several orders of magnitude depending on the area of application within physics, ranging from $\leq 10^{-7}$ for astrophysics and nuclear structure physics, to below $10^{-11}$ for contributions to neutrino mass measurements \cite{Blau2013}. In order to fulfill the demand for higher sensitivity for heavy ions and to improve accuracy for specific studies, a variety of new precision Penning-trap mass spectrometers based on non-destructive detection schemes are at present under commissioning. For example, the PENTATRAP experiment at the Max Planck Institute for Nuclear Physics in Heidelberg has been built to perform ultra-high accuracy mass measurements on highly-charged heavy ions \cite{Repp2012}. There is also a Penning-traps beamline coupled to the TRIGA reactor in Mainz \cite{Kete2008}, and a new Penning-trap system at GSI-Darmstadt, which is part of an upgrade of the SHIPTRAP facility for precision measurements on superheavy elements (SHEs) \cite{Giac2017}. \\


\noindent Nowadays, two detection methods are utilized for precision mass measurements at Radioactive Ion Beam (RIB) facilities: the Time-of-Flight Ion-Cyclotron-Resonance (TOF-ICR) technique, or any of its variants \cite{Koni1995,Geor2007}, and the Phase-Imaging Ion-Cyclotron-Resonance (PI-ICR) technique \cite{Elis2013}. The TOF-ICR technique has been used for several decades, yielding relative mass uncertainties between $10^{-8}$ to $ 10^{-9}$. The PI-ICR technique is a novel method and, by tracking the ion motion using a high-resolution position-sensitive micro-channel plate (MCP) detector, provides a five-fold gain in precision and a 40-fold increase in resolving power. Despite the outcomes obtained by means of these techniques (see e.g. \cite{Bloc2010} for TOF-ICR, and \cite{Elis2015} for PI-ICR), both methods are limited in sensitivity since several tens of ions are needed to obtain a precise motional frequency value. This limits the applicability to the SHEs ($Z\geq 104$) produced in fusion-evaporation reactions with very low production yields, typically of one ion per day or below. The atomic ion with the lowest production yield measured in a Penning trap is $^{256}$Lr$^{++}$ using the TOF-ICR method \cite{Mina2012}. With the PI-ICR technique one could measure a few heavier elements (lower yields) since this method requires smaller number of ions. However, for heavier elements a technique requiring only one ion is mandatory. A well-known technique able to realize this purpose relies on the resonant pick-up of the current a single oscillating ion induces on the trap electrodes. This is the induced-image current technique, which has been used for ultra-precise cyclotron frequency measurements of ions with low or medium mass-to-charge ratios, and of fundamental particles \cite{Corn1989,Haef2003,Vand2006,Stur2014,Ulme2015,Heis2017}, but not demonstrated yet with ions with large mass-to-charge ratios. \\

\noindent Besides the on-going developments of a variant of the induced-image current technique for SHEs \cite{Giac2017,Mina2013}, a novel method for mass spectrometry was proposed based on the use of a single laser-cooled $^{40}$Ca$^+$ ion as detector \cite{Rodr2012}, following a scheme based on two traps described earlier \cite{Hein1990}. On the way towards this experimental realization, the sensitivity of a single $^{40}$Ca$^+$ ion laser-cooled to the Doppler limit ($T _{\hbox{\scriptsize{limit}}}\sim 1$~mK) has been characterized by studying its axial motion in an open-ring Paul trap \cite{Corn2015,Domi2017}. The analytical method and a first approach of using just one trap to store the ion of interest and the laser-cooled ion simultaneously has been recently presented \cite{Domi2017b}, following a former idea described in Ref.~\cite{Drew2004}. Such scheme might allow identifying SHEs elements, after thermalization, transport and capture in a Penning trap with open-ring geometry \cite{Corn2015}, through the fluorescence photons emitted by the $^{40}$Ca$^+$ ion. The identification relies on the determination of  motional frequencies, limiting the relative mass uncertainty $\delta m/m$ in the ion of interest to how precisely one can determine those frequencies from the fluorescence photons. Since the ion of interest is sympathetically cooled by Coulomb interaction with the $^{40}$Ca$^+$ ion, it will be possible to perform precision spectroscopy on a single trapped ion, extending the applicability of the technique used in Ref.~\cite{Laat2016}, to the scenario where the single ion of interest is well under control in a high-intense and homogeneous magnetic field. Furthermore, experiments can be conducted in the quantum-limited regime by using quantum-logic spectroscopy \cite{Schm2005,Ospe2014}.\\

\noindent The open-ring geometry of the Penning trap described in this article differs from all the traps built for precision and accurate mass spectrometry, which are made either by a stack of cylindrical electrodes, see e.g. \cite{Repp2012,Bloc2005}, or with electrodes with the shape of truncated hyperboloids of revolution, as for example the ones described in Refs.~\cite{Mukh2008,Ring2009}. In this publication the full facility around the open-ring Penning trap will be described with special emphasis in the performance of this trap. The first fluorescence measurements obtained with this device will be also presented.

\section{Experimental setup: The Penning-traps beamline}

\begin{figure}[t!]
\hspace{0cm}
\centering\includegraphics[width=1.0\linewidth]{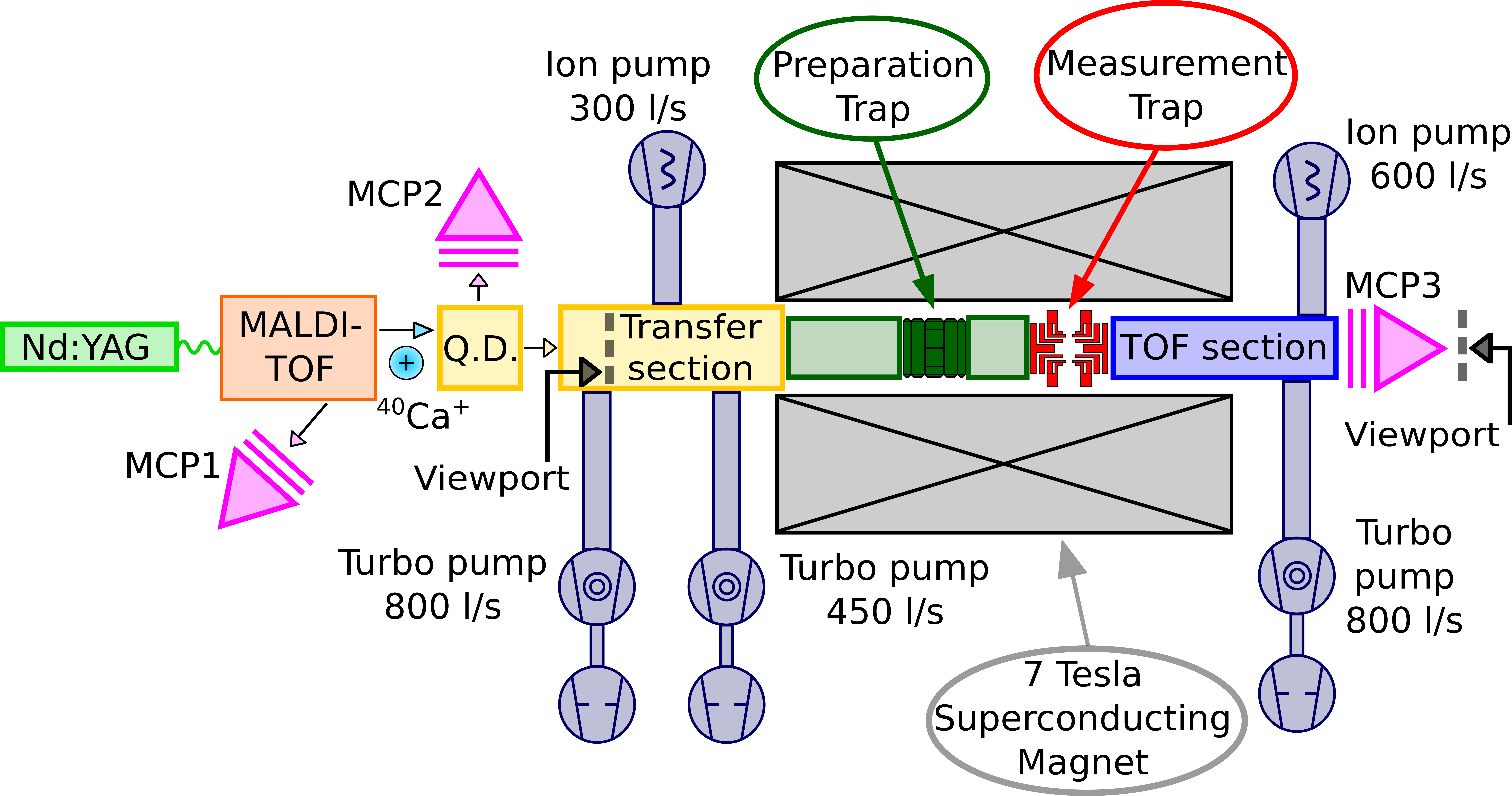}
\caption{Schematic view of the Penning-traps beamline at the TRAPSENSOR laboratory. A view port was placed in the transfer section for the fluorescence measurements, decoupling the MALDI-TOF from the Penning traps. MCP3 is movable in and out of the beamline, for time-of-flight identification and fluorescence measurements, respectively.} \label{Figure2}
\end{figure}

In a real Penning trap built for experiments on ions with mass $m$ and charge $q$, the ions are confined by the superposition of a quadrupolar electrostatic field created by the electrodes, with an homogeneous and strong magnetic field $\vec B$, which defines the axial direction of the revolution symmetry of the device \cite{Brow1986}. The motion of an ion stored in the trap can be depicted as the superposition of three independent modes. The motion along the axial direction does not depend on the magnetic field, has an oscillation amplitude $\rho _z$ and a characteristic frequency given by

\begin{equation}
\nu_z=\frac{1}{2\pi }\sqrt{\frac{q\;U_0}{m\;d^2}},  
\label{eq_axialfrec}
\end{equation}
   
\noindent where in a perfect quadrupolar field, $U_0$ is the potential difference between the endcap and the ring electrode, and $d$ is a parameter defined as

\begin{equation}
d=\sqrt{\frac{1}{2}\left(z_0^2+\frac{1}{2}\rho_0^2\right)}.
\label{Hypd}
\end{equation}
\noindent Here $z_0$ and  $\rho _0$ are the characteristic dimensions of the trap. Assuming the minimum of the trap is at 0~V, the potential the ions experience can be expanded around the trap center, so that
\begin{equation}
U(z)=C_2z^2+C_4z^4+\vartheta (z^6). \label{C4}
\end{equation}
\noindent In an ideal trap, $C_4$ and higher order terms are zero, and Eq.~(\ref{eq_axialfrec}) can be rewritten in terms of Eq.~(\ref{C4}) as
\begin{equation}
\nu_z=\frac{1}{2\pi }\sqrt{\frac{2qC_2}{m}}.  
\label{eq_axialfrec_C2}
\end{equation}
\noindent The other two motional degrees of freedom are in the radial plane, and they are referred to as reduced-cyclotron and magnetron motion, with radii represented by $\rho _+$, and $\rho _-$, respectively, and characteristic frequencies given by
\begin{equation}
\nu_\pm=\frac{\nu_c}{2}\pm\sqrt{\frac{\nu_c^2-2\nu_z^2}{4}}. \label{nu+1}
\end{equation}
Here $\nu _c$ is the cyclotron frequency of an ion in a magnetic field, in the absence of any electrostatic field, and is given by
\begin{equation}
\nu_c=\frac{1}{2\pi }\frac{q}{m}B, \label{cyclotron}
\end{equation}
which relates directly the mass-to-charge ratio of the ion with the observable frequency. In order to perform precise mass measurements, the magnetic field must be strong (on the order of several Tesla) and with an inhomogeneity below the ppm level in the center of the trap across the volume of the ion's motion. The frequency $\nu _c$  can be unfolded from the invariance theorem \cite{Brow1986}
\begin{equation}
\nu_c^2=\nu_+^2+\nu_z^2+\nu_-^2 \label{invariance_theorem}
\end{equation}
\noindent or, in the case of an ideal Penning trap, from the relationship
\begin{equation}
\nu_c=\nu_++\nu_-. \label{tof_identity}
\end{equation}
\noindent  The three eigenfrequencies also follow the relationship $\nu_+\gg \nu_z \gg \nu_-$ so that the determination of the mass through the cyclotron frequency (Eq.~\ref{cyclotron}) implies measuring $\nu_+$ more precisely than the rest of the eigenfrequencies, which is accomplished using ultra-stable magnets. These measurements can be performed using different techniques depending on the accuracy to be reached.\\

\noindent When measurements are envisaged on ions produced at Radioactive Ion Beam facilities, covering a wide kinetic-energy range, more instrumentation has to be added to the Penning trap, in order to stop the ions, to bunch them, and to separate the ion of interest from other ions species also present in the beam but in rates several orders of magnitude larger.\\ 

\noindent Figure~\ref{Figure2} shows a sketch of the TRAPSENSOR Penning-traps beamline built at the University of Granada for laser-based precision experiments comprises a laser-desorption ion source, a transfer section, two Penning traps, one for preparation, including isobaric separation, and one for the measurements, followed by a time-of-flight section. In the following, we describe in detail each of these elements. 

\subsection{Ion production and detection}

\noindent Ions are produced outside the Penning traps by means of a laser-desorption ion source or inside one of the Penning traps via photoionization of an atomic calcium beam. For the fluorescence measurements and the commissioning tests of the setup solely $^{40}$Ca$^+$ ions were used.\\

\noindent The laser-desorption ion source is a commercial apparatus from Bruker Analytical Systems, Model Reflex III, originally designed for
Matrix-Assisted Laser-Desorption-and-Ionization and Time-Of-Flight (MALDI-TOF) identification, to analyze large molecular samples. The source was tested and characterized delivering calcium, osmium and rhenium ions produced at kinetic energies ranging from 20~keV (nominal energy of the apparatus), down to about 100~eV \cite{Corn2013}. For the experiments presented here, the kinetic energy of the ions was fixed to 270~eV. The repetition rate of the frequency-doubled Nd:YAG laser ($\lambda =532$~nm) is 10~Hz, with a pulse length of 4~ns and energies up to 300~mJ. The ion source is coupled to the beamline as shown schematically in Fig.~\ref{Figure2}. The transfer section is made of an electrostatic quadrupole deflector, two steerers and sixteen electrostatic lenses.\\

\noindent Ions are detected and identified by their time-of-flight using one of the several micro-channel plates (MCP) detectors located at different positions along the beamline: MCP1, MCP2 and MCP3 (Fig.~\ref{Figure2}). The MCP3 detector allows performing measurements of the eigenfrequencies as well as of $\nu _c$ of the ions stored in any of the Penning traps after they are ejected towards the detector. 

\subsection{The Penning-traps system}

\begin{figure}[t!]
\hspace{0cm}
\centering\includegraphics[width=1.3\linewidth ]{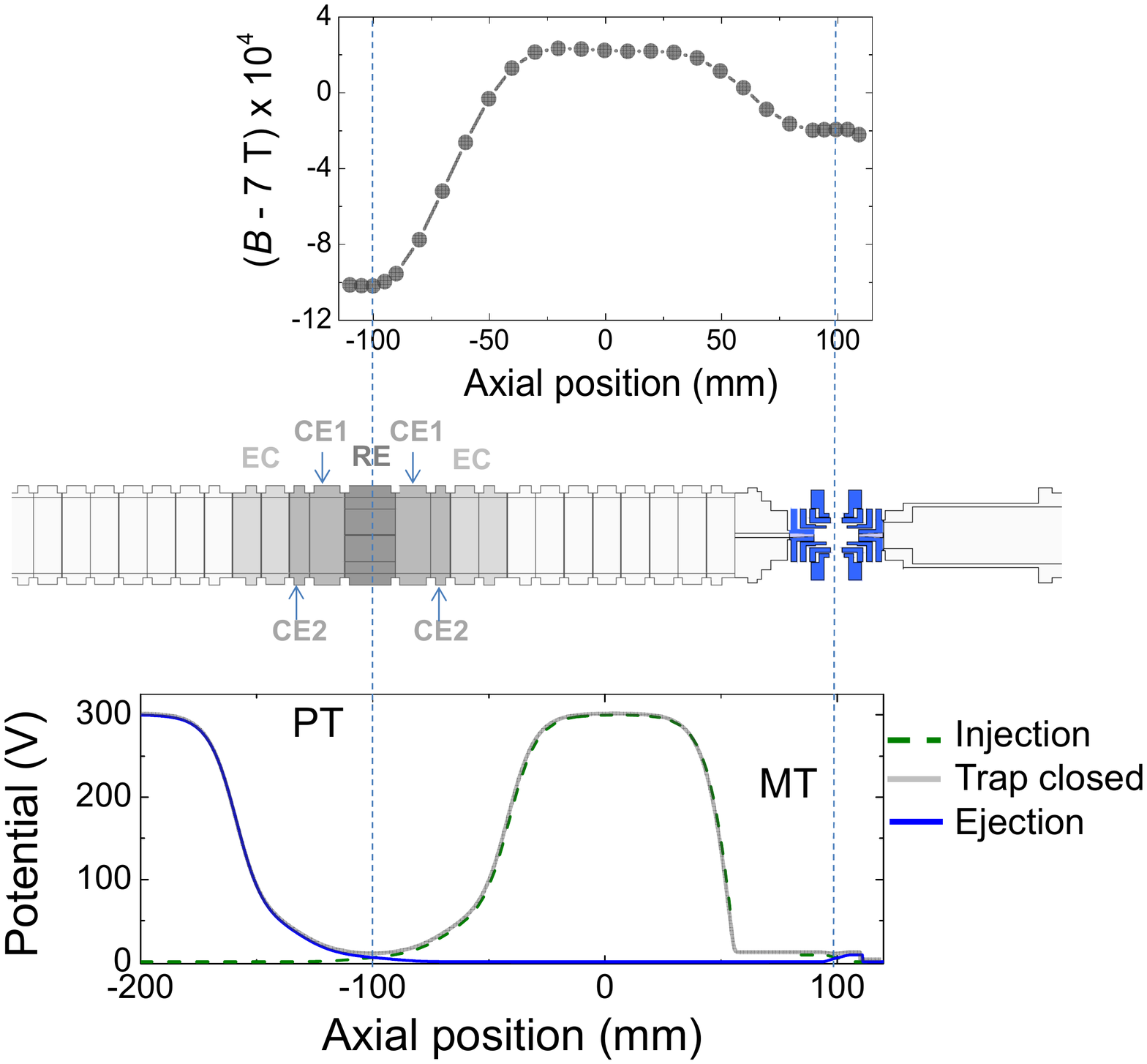}
\vspace{-0.7cm}
\caption{Longitudinal view of the Penning traps (center) showing the magnetic field strength (top) and the potentials along the $z$-axis for injection, capture and extraction (bottom). The acronyms PT and MT stand for Preparation and Measurement Trap, respectively. The electrodes in the center of the PT, i.e. EC (endcap), CE1,2 (correction electrodes) and RE (ring electrode) allows realizing a quadrupolar potential. The potentials of the MT are shown in Fig.~\ref{Figure7}.}\label{Figure3}
\end{figure}

The Penning-traps system comprises two Penning traps housed in the same superconducting solenoid (Fig.~\ref{Figure3}). The superconducting solenoid has the same specifications as those in operation at SHIPTRAP \cite{Bloc2005}, JYFLTRAP \cite{Kolh2004}, MLLTRAP \cite{Webe2013} and TRIGATRAP \cite{Kete2008}. The first trap downstream from the transfer section is made of a stack of 27 cylinders following the conceptual design of the preparation Penning trap for MATS at FAIR \cite{Rodr2010}, so as to allow performing buffer-gas cooling \cite{Sava1991} or electron cooling \cite{Gabr1996}. A longitudinal view of the trap tower together with the magnetic field strength and the potential shape along the $z$-axis, for injection, trapping (and manipulation), and ejection is shown in Fig.~\ref{Figure3}. The center of the trap (grey-shaded area of the CAD drawing) is located in the first homogeneous region of the superconducting solenoid with a magnetic field homogeneity of 10~ppm over 1~cm$^3$. The trap is closed with a potential barrier of 300~V when the ions are delivered with 270~eV kinetic energy. The ring electrode is eight-fold segmented to allow applying dipolar and quadrupolar fields to determine $\nu _+$, $\nu _z$, $\nu _-$ and $\nu _c$. The second trap is an open-ring Penning trap \cite{Corn2015,Corn2016a,Rica2018}, located in a region with a magnetic field homogeneity of 0.14(10)~ppm over the volume of 1~cm$^3$. This trap, coloured in blue and named MT in Fig.~\ref{Figure3}, is used for the measurements. It is described in detail in Sec.~\ref{novel_trap}. For the measurements presented here, the pressure values at both sides of the superconducting solenoid, close to the 800~l/s pumps shown in Fig.~\ref{Figure2}, were around $10^{-8}$~mbar without using the 600~l/s ion pump, located at the end of the beamline, and the 300~l/s one, in front of the solenoid. These pumps are required for the vacuum level in order to perform fluorescence measurements.\\

\noindent The PT was designed to allow cooling ions via collisions with gas atoms, but also to implement other cooling mechanisms compatible with ultra-high vacuum, since the pumping barrier between the traps (with a length of 23~mm and a diameter of 2~mm), limits the pressure difference between the PT and the MT. Thus, buffer-gas cooling in the PT restricts those experiments based on creating a two-ion crystal in the MT as envisaged \cite{Domi2017b}. However, cooling of ions in the PT should be possible by their interaction with a bath of electrons trapped at room temperature in a nested trap \cite{Gabr1996}, or with a cloud of laser-cooled ions, as proposed in Ref.~\cite{Buss2006}, and demonstrated later by sympathetically cooling highly-charged ions in a Paul and in a Penning trap \cite{Schm2015,Schm2017}. 

\subsubsection{Cooling and mass separation}

\begin{figure}[t!]
\hspace{0cm}
\centering\includegraphics[width=1.0\linewidth ]{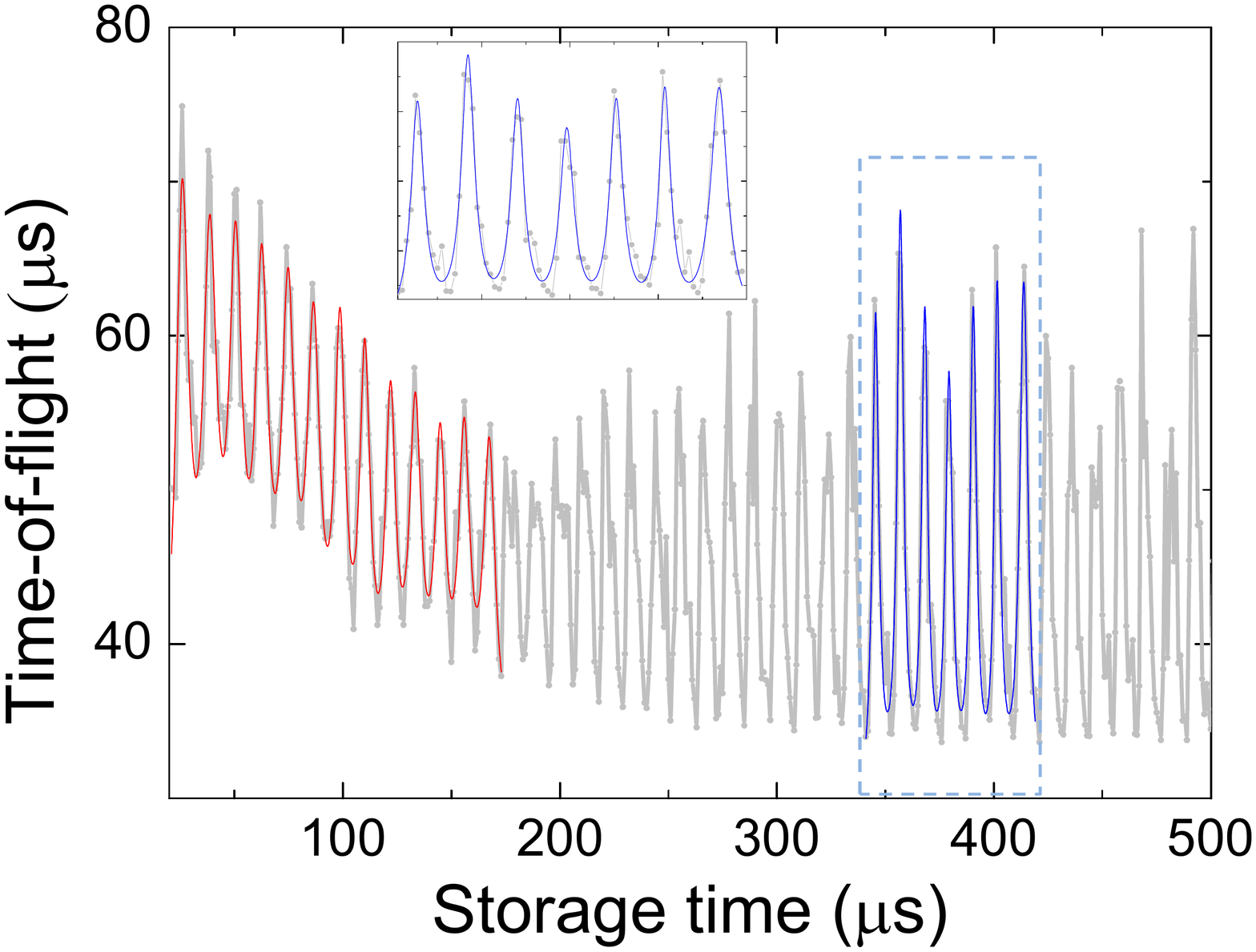}
\vspace{-1.2cm}
\caption{Time-of-flight of $^{40}$Ca$^+$ ions from the PT to MCP3 (see Fig.~\ref{Figure2}) versus storage time in the PT for times below 500~$\mu$s. One of the correction electrodes at the injection side of the trap was connected to a circuit for induced current detection, and is polarized through a low-pass filter. The filter causes the rise time during ion capture to be slower for that correction electrode, and is thus responsible for the higher time-of-flight for storage times below 100~$\mu$s. The blue and red solid-lines are the results from pulti-peak fits.}\label{Figure4}
\end{figure}

\begin{figure}[t!]
\hspace{0cm}
\centering\includegraphics[width=1.0\linewidth ]{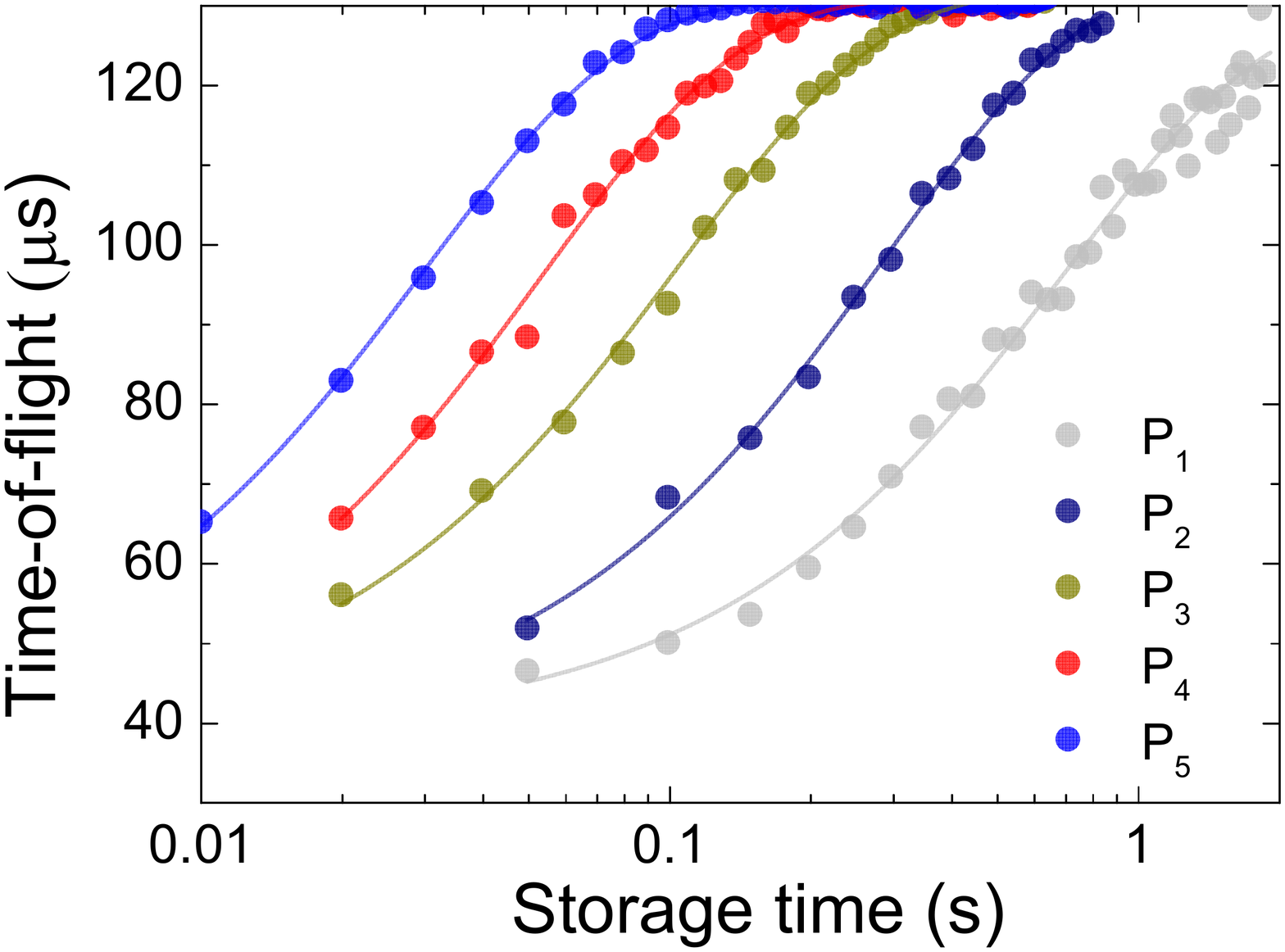}
\vspace{-1.3cm}
\caption{Time-of-flight of $^{40}$Ca$^+$ ions from the PT to MCP3 (see Fig.~\ref{Figure2}) as a function of the storage time in the PT for different buffer-gas pressures. The solid lines represent fits using the exponential grow function of Eq.~(\ref{exponential}). The pressure values inside the PT are obtained from the fit yielding $p_1=1.3\times 10^{-6}$~mbar ($\tau _z^{\hbox{\scriptsize{cool}}}\approx 700$~ms), $p_2=3.0\times 10^{-6}$~mbar ($\tau _z^{\hbox{\scriptsize{cool}}}\approx 300$~ms), $p_3 = 8.3\times 10^{-6}$~mbar ($\tau _z^{\hbox{\scriptsize{cool}}}\approx 100$~ms), $p_4 = 1.7\times 10^{-5}$~mbar ($\tau _z^{\hbox{\scriptsize{cool}}}\approx 50$~ms), and $p_5=2.9\times 10^{-5}$~mbar ($\tau _z^{\hbox{\scriptsize{cool}}}\approx 30$~ms).}\label{Figure5}
\end{figure}

\noindent The ions stored in the harmonic potential of the trap with axial frequency $\omega_{z}=2\pi \nu _z$ experience a restoring force of the form $\vec F=-2m\delta_{z}\vec v$, where $m$ is the ions' mass and $\vec v$ their velocity. The cooling rate $\gamma_{\rm{z}}=2\delta_{z}$ of the ions' kinetic energy along the $z$-direction is determined by 

\begin{equation}
\delta_{\rm{z}}=\frac{q}{2m}\frac{1}{\mu_{0}}\frac{p_{\rm{b}}/p_{0}}{T_{\rm{b}}/T_{0}},
\label{Eq:GammeZ}
\end{equation}

\noindent with the ion mobility $\mu_{0}$, the normalized residual gas pressure $p_{\rm{b}}/p_{0}$ and the normalized buffer-gas temperature $T_{\rm{b}}/T_{0}$. $p_{0}$ and $T_{0}$ are 1013\,mbar and 293\,K, respectively. The thermal ion mobility of singly-charged $^{40}$Ca$^+$ ions in a helium buffer-gas is considered equal to that of Ar$^+$ ions such that $\mu _0=21.2\times 10^{-4}$\,m$^2$s$^{-1}$/V \cite{Elli1984}. The corresponding cooling time constant is $\tau _z^{\hbox{\scriptsize{cool}}}=1/\gamma _z$. Figs.~\ref{Figure4} and \ref{Figure5} show the time-of-flight of $^{40}$Ca$^+$ ions ejected from the PT towards the detector MCP3, as a function of the storage time. In Fig.~\ref{Figure4}, one can see oscillations in the time-of-flight spectrum, depicting the axial oscillation of the ions in the PT during the first 500~$\mu $s of the cooling process. The axial oscillation frequency of the $^{40}$Ca$^+$ ions within this process varies from $\nu _z = 40.9(3)$~kHz, within the first hundred microseconds, to $\nu _z = 43.6(2)$~kHz for a storage time around 500~$\mu $s. These frequency values are below those obtained from the measurements of $\nu _{\pm}$ and $\nu _c$, after the ions are cooled during a few hundreds of milliseconds. The eigenfrequencies for $^{40}$Ca$^+$ are shown in Tab.~\ref{eigenfrequencies_PT}. The radial eigenfrequencies were measured by probing the ions' motion in the radial direction with external dipolar fields varying $\nu _{\hbox{\scriptsize{RF}}}$ around $\nu _{\pm}$.\\

\begin{table}[b!]
\caption{Eigenfrequencies for $^{40}$Ca$^+$ when the PT (Fig.~\ref{Figure3}) is operated with $V_{\hbox{\scriptsize{EC}}}=300$~V, $V_{\hbox{\scriptsize{CE1}}}=45$~V, $V_{\hbox{\scriptsize{CE2}}}=29.5$~V, and $V_{\hbox{\scriptsize{RE}}}=5$~V. ($^*$) This value was obtained using the measured values of $\nu _c$, and $\nu _{\pm}$ together with Eq.~(\ref{invariance_theorem}).}
\vspace{0.2cm}
\centering
 \renewcommand{\arraystretch}{1.3}
\setlength{\tabcolsep}{7.mm}
\label{Tab:1}
	\begin{tabular}{ccc}
			\hline\hline
 $\nu _+$ (MHz) & $\nu_z$  (kHz) & $\nu _-$  (Hz) \\

\hline \hline
			$2.688592(2)$ &  49.500(140)$^*$& 456(7)\\

	\hline \hline 
	\end{tabular} \label{eigenfrequencies_PT}
\end{table}

\noindent The effect of the buffer-gas pressure on the cooling time is shown in Fig.~\ref{Figure5}. The data points are fitted using the storage-time dependent function given by

\begin{equation}
S(t)=S_M-(S_M-S_m)e^{-t/\tau _z^{\hbox{\scriptsize{cool}}}}, \label{exponential}
\end{equation}

\noindent where $S$ represents here the average time-of-flight signal, and the subscripts $M$ and $m$ stand for the maximum and the minimum time-of-flight values, respectively. The helium buffer-gas pressure inside the PT is obtained from $\tau^{\hbox{\scriptsize{cool}}}$ using Eq.~(\ref{Eq:GammeZ}).\\

\noindent In case of the radial motions $\omega_{\pm}$ the damping constants read
\begin{equation}
\delta_{\rm{\pm}}=2\delta_{\rm{z}}\frac{\omega_{\pm}}{\omega_{+}-\omega_{-}}.
\end{equation}

\noindent Thus, for a pressure of $1.3\times 10^{-6}$~mbar ($\tau _z^{\hbox{\scriptsize{cool}}}\approx 700$~ms), $\delta _z \approx 0.7$~s$^{-1}$, $\delta _+ \approx 1.4$~s$^{-1}$, and $\delta _-\approx 2.4\times 10^{-4}$~s$^{-1}$. This yields a decay time constant for the magnetron motion of $\tau _-^{\hbox{\scriptsize{cool}}}\approx 4100$~s, and thus, an increase of the magnetron radius within the storage time of the ions inside the PT is negligible.\\ 

\begin{figure}[t!]
\hspace{0cm}
\centering\includegraphics[width=1.0\linewidth ]{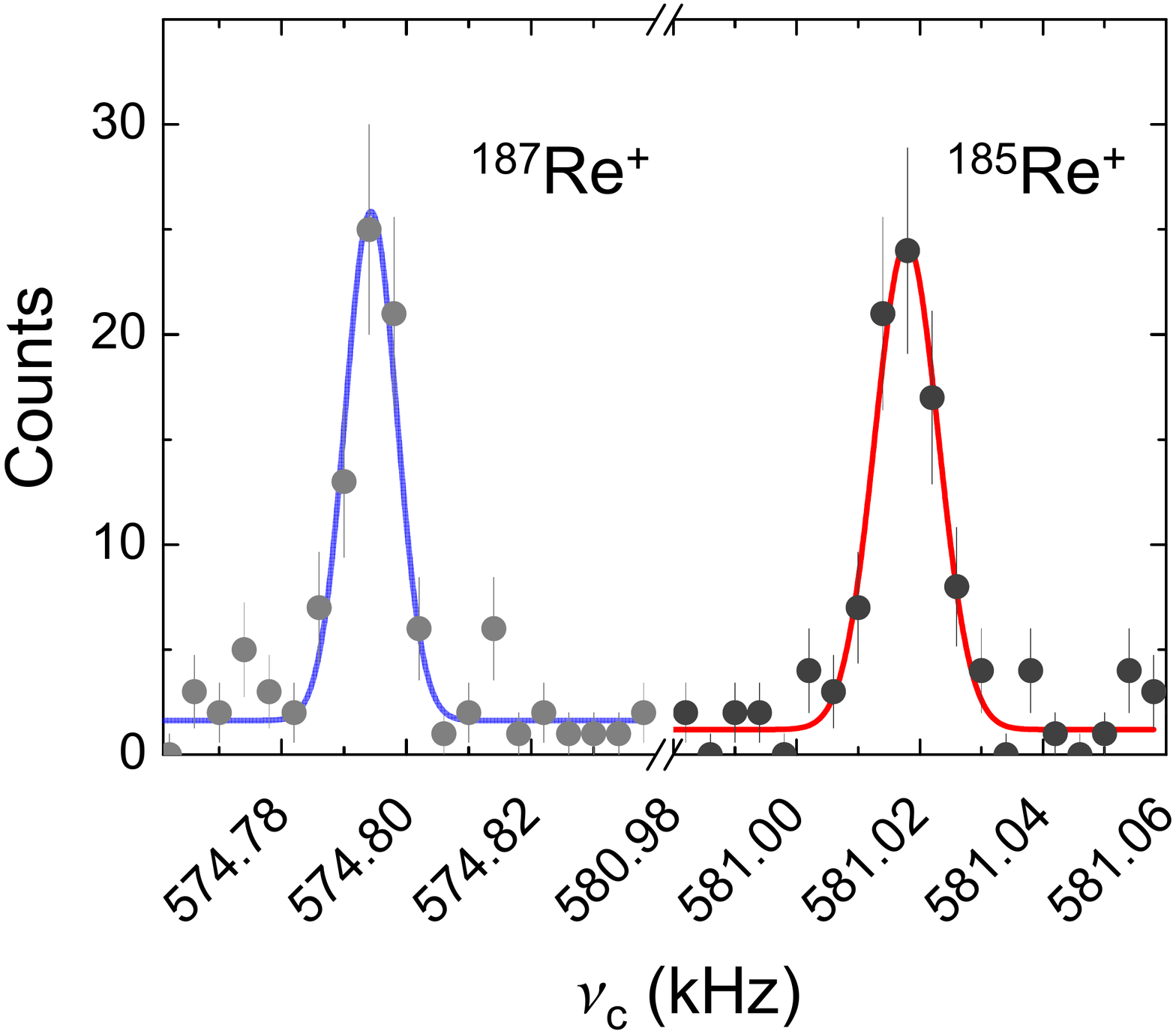}
\vspace{-1.5cm}
\caption{Cooling resonances in the PT for $^{185,187}$Re$^+$ ions. The resolving power $m/\Delta m$ is $5.7\times 10^4$ and $4.8\times 10^4$ for $^{187}$Re$^+$ and $^{185}$Re$^+$, respectively.}\label{Figure6}
\end{figure}

\noindent The main purpose of the PT, which is centering the ions with specific mass-to-charge ratio to allow removing unwanted species, was tested using $^{40}$Ca$^+$ and $^{185,187}$Re$^+$ ions. The performance of the device with respect to the buffer-gas cooling technique \cite{Sava1991} was described in Refs.~\cite{Corn2016b,Corn2016c}. This method allows centering the ions with a specific mass-to-charge ratio when combining the collisions with buffer-gas atoms, with the application of an external quadrupolar field in the radial direction at $\nu _{\hbox{\scriptsize{RF}}}=\nu _c$. Figure~\ref{Figure6} shows cooling resonances for the two naturally abundant rhenium isotopes, $^{185}$Re and $^{187}$Re. The frequencies for $^{40}$Ca$^+$ and $^{185,187}$Re$^+$ ions and the resulting resolving powers are listed in Tab.~(\ref{cyclotron_PT}). \\

\begin{table}[b!]
\caption{Cyclotron frequency ($\nu _c$) and resolving power ($m/\Delta m$) from cooling resonances in the PT.}
\vspace{0.2cm}
\centering
 \renewcommand{\arraystretch}{1.3}
\setlength{\tabcolsep}{7.mm}
\label{Tab:1}
	\begin{tabular}{ccc}
			\hline\hline
 Ion species & $\nu_c$  (Hz) & $m/\Delta m$ \\

\hline \hline
			 $^{40}$Ca$^+$ &  $2\,689\,067.5(3)$&  $2.7\times 10^5$\\
 			$^{185}$Re$^+$ &  581\,017.9(7)& $4.8\times 10^4$\\
			 $^{187}$Re$^+$&  574\,794.5(5)& $5.7\times 10^4$\\

	\hline \hline 
	\end{tabular}\label{cyclotron_PT}
\end{table}

\noindent The laser-desorption ion source delivers ions with 10-ns duration pulses directly into the PT every 100~ms. Since a measurement cycle might last up to several tens of seconds, stacking has been also implemented. 

\section{The open-ring Penning trap}\label{novel_trap}

\begin{figure}[t!]
\hspace{0cm}
\centering\includegraphics[width=1.0\linewidth ]{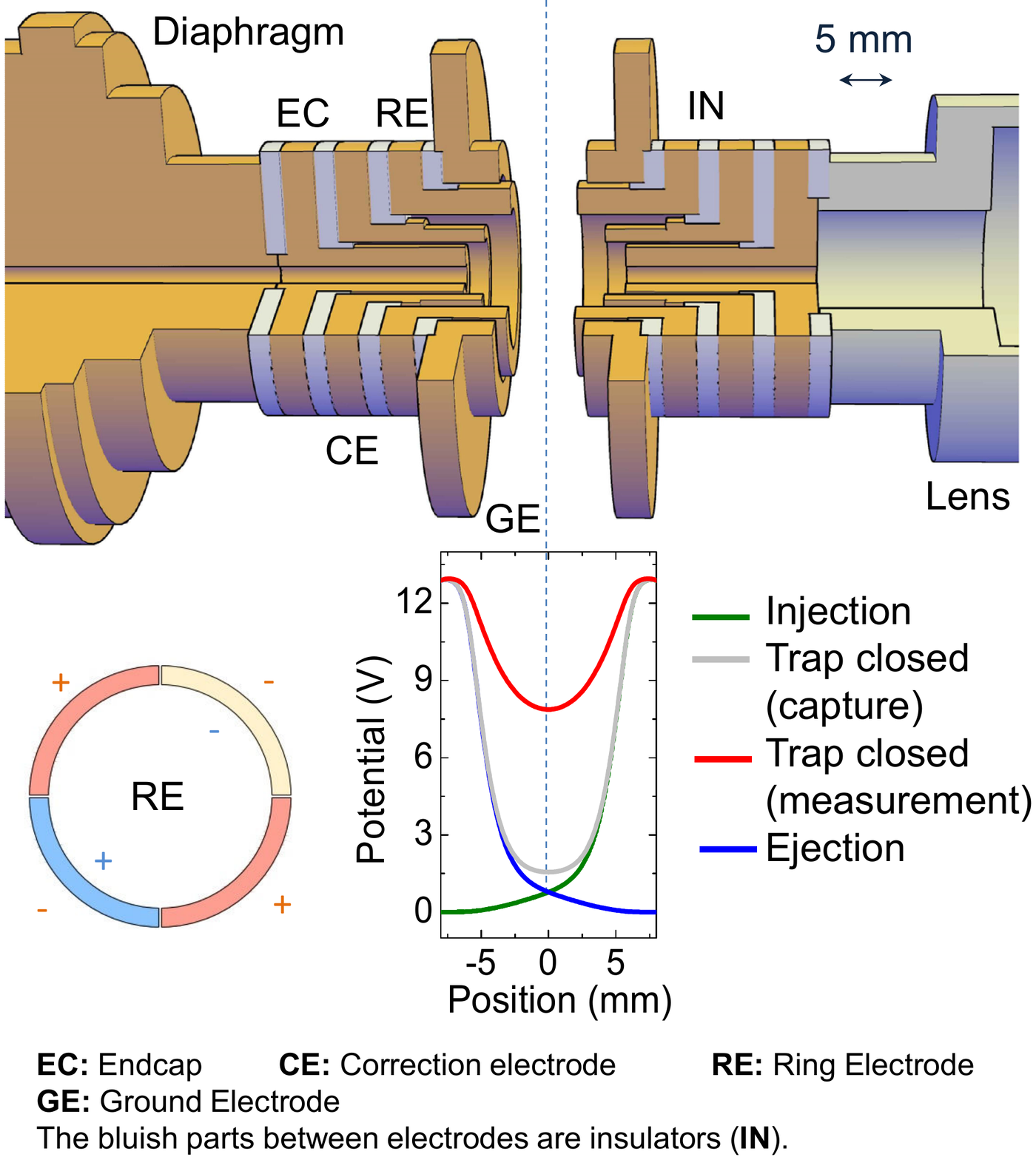}
\vspace{-1.0cm}
\caption{Three-dimensional CAD drawing of the open-ring Penning trap together with the potentials applied for injection, capture, measurement, and extraction. The ring electrode (RE) is four-fold segmented to allow applying dipolar or quadrupolar RF excitations.}\label{Figure7}
\end{figure}

The measurement Penning trap (MT) is an open-ring trap, shown in detail in Fig.~\ref{Figure7}, together with the trap potential for injection, trapping and ejection of the ions. The trap is made of two sets of four concentric rings forming the endcap electrodes (EC), the ring electrodes (RE), corrections electrodes (CE) and an external electrode, named grounded electrode (GE). The trap is located between the diaphragm, built after the PT, and the first electrostatic lens in the time-of-flight section. The ring electrodes are four-fold segmented in order to apply external radiofrequency fields with different polarities (dipolar and quadrupolar) in order to probe the eigenfrequencies of the ions and $\nu _c$.

\subsection{Trap tuning}

\begin{figure}[t!]
\hspace{0cm}
\centering\includegraphics[width=1.0\linewidth ]{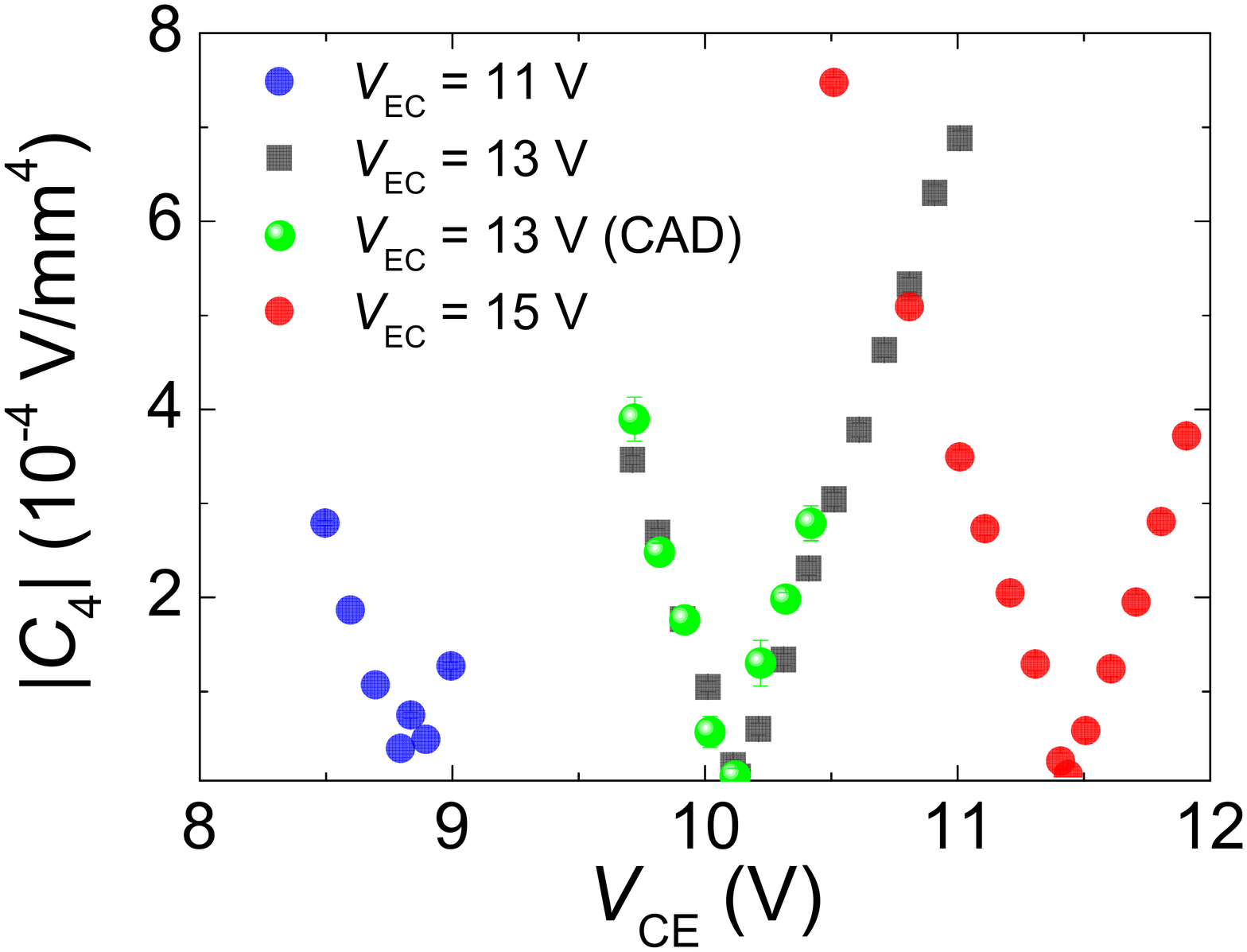}
\vspace{-1.5cm}
\caption{Absolute value of the coefficient $C_4$ as a function of the voltage applied to the correction electrode ($V_{\scriptsize{\hbox{CE}}}$). The data points are obtained after fitting with the function made out of the first two terms in Eq.~(\ref{C4}), the potential shape along the $z$-axis obtained from SIMION simulations. For this purpose the CAD file shown in Fig.~\ref{Figure7} was converted to a SIMION .PA$\#$ file using SL Tools. }\label{Figure8}
\end{figure}

The geometry of the trap (Fig.~\ref{Figure7}) was shown previously in Ref.~\cite{Corn2015}, introducing this trap for laser cooling experiments. The device was similar to the one built at GANIL (LPCTRAP) \cite{Rodr2006} for $\beta$-$\nu$ correlation experiments \cite{Flec2008,Flec2011} and in both cases was run as a Paul trap. The Penning trap presented here has been scaled down by a factor of $2$. The quadrupolarity of the potential has been investigated through SIMION simulations \cite{SIMION} for different voltage configurations in order to minimize the coefficient $C_4$ (Eq.~(\ref{C4})). The potential along the $z$-axis is recorded and fitted utilizing the function given by the first two terms in Eq.~(\ref{C4}). Figure~\ref{Figure8} shows $|C_4|$ as a function of the voltage applied to the correction electrodes $V_{\scriptsize{\hbox{CE}}}$, for three different values of $V_{\scriptsize{\hbox{EC}}}$. For all the data points $V_{\scriptsize{\hbox{RE}}}=5.8$~V and  $V_{\scriptsize{\hbox{GE}}}=0$~V. A similar procedure was followed (the results are not shown) keeping a constant value of $V_{\scriptsize{\hbox{EC}}}$, equal to 13~V. In this case, $C_4$ was calculated as a function of $V_{\scriptsize{\hbox{CE}}}$ for three different values of $V_{\scriptsize{\hbox{RE}}}$, namely, 1.8, 3.8, and 5.8~V. $V_{\scriptsize{\hbox{GE}}}$ was set to $0$~V. From the results obtained using SIMION, $C_4$ can be minimized by taking
\begin{equation}
V_{\scriptsize{\hbox{CE}}}=0.655(3)\cdot V_{\scriptsize{\hbox{EC}}}+1.63(4), \label{tuning1}
\end{equation}
\noindent or $V_{\scriptsize{\hbox{CE}}}=0.2736(11)\cdot V_{\scriptsize{\hbox{RE}}}+8.552(4)$, when tuning the trap by varying the voltage applied to the ring electrode as a function of $V_{\scriptsize{\hbox{EC}}}$ or $V_{\scriptsize{\hbox{RE}}}$, respectively. $\nu _z$ is obtained for each configuration from the coefficient $C_2$ (Eq.~(\ref{eq_axialfrec_C2})). For the voltages applied, $\nu _z$ was varied from about $100$ to $200$~kHz. Using again the results from the SIMION simulations (keeping $V_{\scriptsize{\hbox{RE}}}=5.8$~V), it is possible to build the polynomial function

\begin{equation}
\frac{V_{\scriptsize{\hbox{EC}}}}{V_{\scriptsize{\hbox{CE}}}}= A_0+A_1\nu _z +A_2 \nu _z^2, \label{tuning2}
\end{equation}

\noindent with $A_0=0.873(21)$,  $A_1=0.0040(3)$/kHz, and $A_2=-8(1)\times 10^{-6}$/(kHz)$^2$.  Deciding on an axial frequency, the voltages to be applied at EC and CE are obtained by solving the system formed by Eqs.~(\ref{tuning1}) and (\ref{tuning2}). This allows tuning the trap close to the final value of the experiment. Note that the power supply (model BS1-8 from Stahl Electronics) delivers signals in the range $\pm15.000(1)$~V. By changing $V_{\scriptsize{\hbox{EC}}}$ and $V_{\scriptsize{\hbox{CE}}}$, the bottom of the potential along the MT is biased from 6 to 9~V, so that the ions can be captured using the scheme shown in Fig.~\ref{Figure7}. 

\begin{figure}[t]
\hspace{0cm}
\centering\includegraphics[width=1.0\linewidth ]{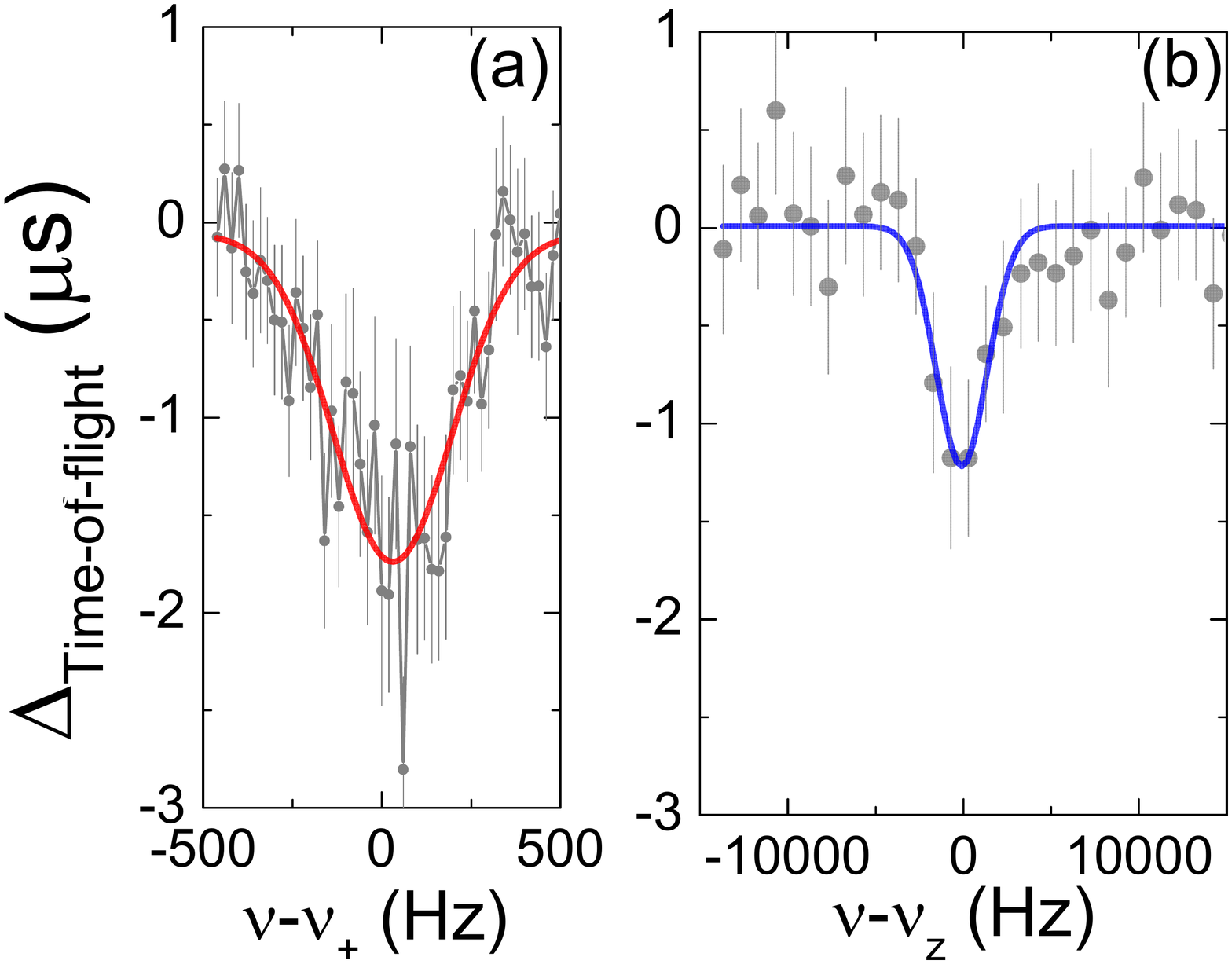}
\vspace{-0.6cm}
\caption{(a) Determination of the radial eigenfrequencies of $^{40}$Ca$^+$ ions in the open-ring trap. (b) Determination of the axial frequency (Eq.~(\ref{eq_axialfrec})) of $^{40}$Ca$^+$ ions in the open-ring trap.The excitation time was 100~ms.}\label{Figure9}
\end{figure}

\subsection{Trap performance}

The performance of the trap has been investigated from measurements of $\nu_+$, $\nu _z$, $\nu _-$, and $\nu _c$, and simulations of the time-of-flight signal of $^{40}$Ca$^+$ ions from the MT to the MCP3 through the TOF section. The average time-of-flight of the ions from the center of the MT ($z_{\scriptsize{\hbox{MT}}}$) to the detector MCP3 ($z_{\hbox{\scriptsize{det}}}$) is given by \cite{Koni1995}
\begin{equation}
S=\int_{0}^{z_{\hbox{\scriptsize{det}}}} \sqrt{\frac{m}{2(E_{z}(t)-qU(z)+|\mu(E_{r}(t))B(z)|)}} \, \rm{d}z,
\label{Eq:TOF}
\end{equation}
\noindent where $E_{z}(t)$  and $E_{r}(t)$ are the ion's axial  and radial energy, respectively, as a function of time $t$, and $U(z)$ and $B(z)$ are the electric potential and magnetic field strength along the $z$-axis, respectively. The ion's orbital magnetic moment $\mu$ is given by 
\begin{equation}
|\mu|=|\pi\cdot q\cdot (\rho_{+}^2\nu_{+}+\rho_{-}^2\nu_{-})|=\left |\frac{E_{r}}{B_{0}}\right |.
\label{Eq:magneticMoment}
\end{equation}
\begin{table}[b!]
\caption{Eigenfrequencies for $^{40}$Ca$^+$, determined by measuring the time-of-flight of the ions from the MT to the MCP3 detector, after probing the eigenmotions with an external dipolar field (Fig.~\ref{Figure9}). The MT (Fig.~\ref{Figure7}) is operated with $V_{\hbox{\scriptsize{EC}}}=13$~V, $V_{\hbox{\scriptsize{RE}}}=5.8$~V, and $V_{\hbox{\scriptsize{GE}}}=0$~V. The table also lists the values of $U_0/d^2$ obtained from $\nu _+$ (Eq.~(\ref{nu+})) and $\nu _z$ (Eq.~(\ref{eq_axialfrec})).}
\vspace{0.2cm}
\centering
 \renewcommand{\arraystretch}{1.3}
\setlength{\tabcolsep}{1.2mm}
\label{Tab:1}
	\begin{tabular}{ccccc}
			\hline\hline
$V_{\hbox{\scriptsize{CE}}}$ &  $\nu _+$  & $U_0/d^2$& $\nu_z$  & $U_0/d^2$\\
(V)&  (MHz) & V/mm$^2$& (kHz)& V/mm$^2$\\

\hline \hline
			9.82 & 2.685203(11)&  0.3667(12)&149.916(93)& 0.3674(5)  \\
                              10.14 & 2.685135(17)& 0.3727(18) &150.277(231)& 0.3692(11)  \\
                              10.42 & 2.685090(27)&  0.377(3)&152.109(679)& 0.378(4)\\
	\hline \hline 
	\end{tabular} \label{eigenfrequencies1}
\end{table}





\noindent Figure~\ref{Figure9}(a) shows a measurement of the reduced-cyclotron frequency obtained by applying an external dipolar field at $\nu _+$ via the ring electrode segments. In the absence of any external field, the mean time-of-flight of the extracted $^{40}$Ca$^+$ ions from the MT to the MCP3 detector is 75.5(1)~$\mu$s, which corresponds to an initial ions' mean kinetic energy of 2.08(6)~eV. This is the outcome after comparing the experimental results with those obtained from SIMION simulations by implementing the MT together with the lenses in the time-of-flight section, the MCP3 detector and the magnetic field gradient. Furthermore, using a different voltage configuration during the ejection of the ions, the initial mean kinetic energy is also $\approx 2.1$~eV, for a measured average time-of-flight of 78.9(3)~$\mu$s. From cooling measurements in the MT, performed in similar way as those shown in Fig.~\ref{Figure5}, the ions are thermalized in the trap, so that this energy might be gained in the extraction process, or there should be a time-of-flight offset arising from the signal processing and/or the acquisition system. Assuming this energy is only axial energy, the radii of the reduced-cyclotron orbits the ions would have to yield the minimum time-of-flights in Fig.~\ref{Figure9}(a) are $\rho _+\approx 250$~$\mu$m (left) and $\rho _+\approx 160$~$\mu$m (right). The time-of-flight effect after probing the axial motion is shown in Fig.~\ref{Figure9}(b). The mean kinetic energy of the ions is 3.8~eV. However, since the initial energy is biased, the energy originated from the external field is $1.7$~eV, corresponding to a maximum axial oscillation amplitude $\rho _z =4.6$~mm.\\


\noindent The eigenfrequencies of $^{40}$Ca$^+$ ions are shown in Tab.~\ref{eigenfrequencies1}. The voltage applied to the CE was around the minimum shown in Fig.~\ref{Figure8} when $V_{\hbox{\scriptsize{EC}}}=13$~V. These frequencies are obtained from the time-of-flight of the ions from the MT to the detector after probing their eigenmotions. From the measured values of $\nu _+$ and $\nu _z$, it is possible to obtain the ratio $U_0/d^2$ from


\begin{equation}
\nu_+=\frac{qB}{2\pi m}-\frac{U_0}{4\pi d^2B} \label{nu+}
\end{equation}
\noindent and Eq.~(\ref{eq_axialfrec}). 


\subsection{Cyclotron-frequency measurements}

\begin{figure}[t]
\hspace{0cm}
\centering\includegraphics[width=1.0\linewidth ]{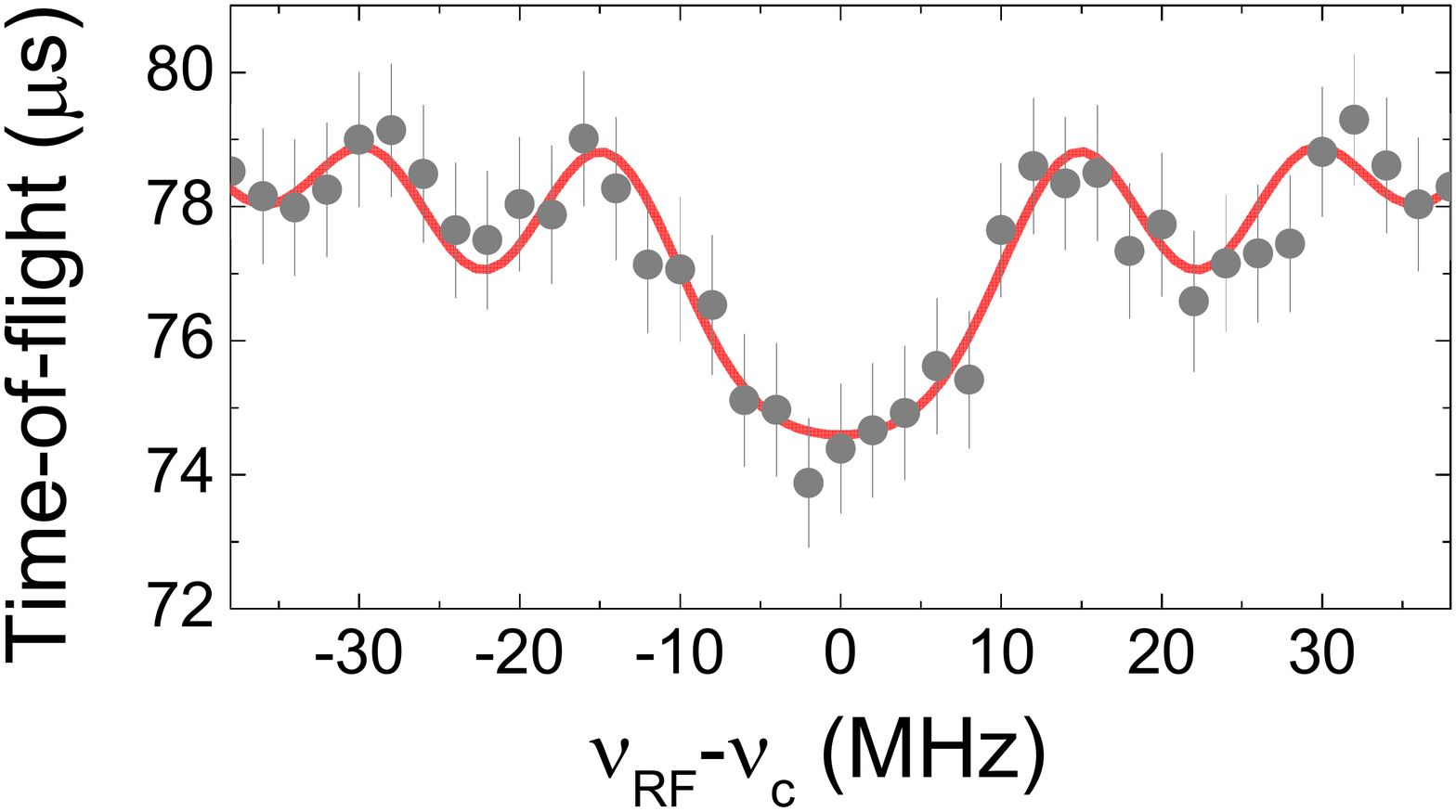}
\vspace{-3cm}
\caption{Time-Of-Flight Ion-Cyclotron-Resonance for $^{40}$Ca+ ions. The voltage configuration was Config.~1 (Tab.~\ref{nu_cyclotron}). $\nu _c$ is the cyclotron frequency. See text for further details.}\label{Figure11}
\end{figure}

\begin{table}[b!]
\caption{MT cyclotron frequency measurements for $^{40}$Ca$^+$ for different voltage configurations (Config.). $V_{\hbox{\scriptsize{CE}}}=9.82$~V and $V_{\hbox{\scriptsize{GE}}}=13$~V in Config.~1, and   $V_{\hbox{\scriptsize{CE}}}=10.14$~V and $V_{\hbox{\scriptsize{GE}}}=0$~V in Config.~2. $V_{\hbox{\scriptsize{EC}}}=13$~V, and $V_{\hbox{\scriptsize{RE}}}=5.8$~V in both configurations. The values of $\nu _c$ (TOF-ICR) are the mean values from seven measurements. The uncertainties correspond to 1~$\sigma$. ($^{\#}$) $\nu _z =146.918(208)$~kHz was obtained using $2\nu _+\nu_-=\nu _z ^2$, and the measured values of $\nu _+ =2.685360(7)$~MHz and $\nu _- =4019(6)$~Hz. $|C_4^{\hbox{\scriptsize{ratio}}}|$ is the absolute value of $C_4$ normalized to $C_4$ for Config.~2. }
\vspace{0.2cm}
\centering
 \renewcommand{\arraystretch}{1.3}
\setlength{\tabcolsep}{1.0mm}

	\begin{tabular}{ccccc}
			\hline\hline
Config.& $\nu _c$   (TOF-ICR) &  $\nu _c$ (Eq.~\ref{invariance_theorem})& $\nu _c$  (Eq.~\ref{tof_identity})& $|C_4^{\hbox{\scriptsize{ratio}}}|$\\

 &  (MHz) &  (MHz) & (MHz) &  \\

\hline \hline
			1 &  2.6893743(7)& 2.689379(19)$^{\#}$& 2.689379(13)& $4.3$\\
                      	2&  2.6893723(9)& 2.689340(30)& 2.689428(35)&$1$\\

	\hline \hline 
	\end{tabular} \label{nu_cyclotron}
\end{table}

The mass-to-charge ratio of an ion is determined from its cyclotron frequency using Eq.~(\ref{cyclotron}), where the magnetic-field strength $B$ is measured via the cyclotron frequency of an reference ion with a precisely known mass value (see e.g. \cite{Blau2013}). In turn, $\nu _c$ can be unfolded from the measurements of the eigenfrequencies of the ions using Eq.~(\ref{invariance_theorem}), or by applying an external quadrupolar radiofrequency field that will originate energy exchange between the radial eigenmodes (see Eq.~(\ref{tof_identity})). Scanning the radiofrequency $\nu _{\hbox{\scriptsize{RF}}}$ around $\nu _c$ and measuring, for each frequency value, the time-of-flight of the ions from the MT to the detector, results in a TOF-ICR measurement \cite{Koni1995}, similar as the one shown in Fig.~\ref{Figure11} for $^{40}$Ca$^+$ ions. In this case, a radiofrequency field with an amplitude $V_{\hbox{\scriptsize{RF}}}=76$~mV$_{\hbox{\scriptsize{pp}}}$ was applied for a time $T_{\hbox{\scriptsize{RF}}}=$100~ms. This amplitude drives five conversions between magnetron and reduced-cyclotron motion.  More TOF-ICR measurements using $^{40}$Ca$^+$ ions have been taken for $T_{\hbox{\scriptsize{RF}}}=50$, 100 and 200~ms. The maximum mass resolving power obtained for $^{40}$Ca$^+$ from the TOF-ICR measurements is $m/\Delta m =6.6\times 10^{5}$. A single TOF-ICR measurement was taken to determine $\nu _c$  for $^{187}$Re$^+$, yielding 574.8861(25)~kHz ($T_{\hbox{\scriptsize{RF}}}=25$~ms and $m/\Delta m =1.8\times 10^{4}$).\\

\noindent Equation~(\ref{invariance_theorem}) is utilized to obtain $\nu _c$ when the eigenfrequencies are measured using a non-destructive detection technique, i.e. detecting the current induced by a stored ion in the trap electrodes (e.g. \cite{Corn1989}) or, as envisaged in our future experiments, using fluorescence photons from a laser-cooled $^{40}$Ca$^+$ ion \cite{Rodr2012}. However, in this publication we have measured the eigenfrequencies via time-of-flight, after applying an external dipolar field. The three cyclotron-frequency values obtained from the eigenfrequencies listed in Tab.~(\ref{eigenfrequencies1}), are 2.689388(16), 2.689340(30), and 2.689342(65)~MHz, which are in agreement with $\nu _c=2.689373(3)$~MHz, resulting from 30 measurements carried out by means of the TOF-ICR technique \cite{Koni1995}, during a period of six months, after performing the regular $B_0$-dump maintenance of the magnet. For these measurements, several voltage configurations were adopted yielding a standard deviation in $\nu_c $ of $\approx 2.7$~Hz, and thus ensuring the harmonicity of the field. The relative fluctuations of the magnetic field have been recently measured during one hour, obtaining $\delta B /B =1.9\times 10^{-7}$.

\section{Fluorescence measurements: towards Doppler cooling in 7 Tesla}

The main scientific goal of the TRAPSENSOR Penning-traps facility is to use a single laser-cooled $^{40}$Ca$^+$ ion as high-sensitive sensor for precision Penning-trap mass spectrometry \cite{Rodr2012}. The characterization of a single laser-cooled $^{40}$Ca$^+$ ion was accomplished using a Paul trap \cite{Domi2017}, including the scenario with two ions in the trap \cite{Domi2017b}. From the technical point of view, moving the experiment from a Paul-trap platform to another one based on a 7-T Penning trap complicates the setup due to the Zeeman effect on the $^{40}$Ca$^+$ atomic transitions and the complex image collection system. A detailed description of the gain in sensitivity and precision using the single laser-cooled $^{40}$Ca$^+$ ion in a 7 Tesla Penning trap will be presented elsewhere.

\subsection{Level scheme, frequency stabilization and laser beam transport}

$^{40}$Ca$^{+}$ is alkali-like, and thus it has a single valence electron in the $4^{2}$S$_{1/2}$ ground state. The first excited state is the $4^{2}$P$_{1/2}$-state, which decays with a probability of 12/13 back into the ground state (at $\lambda = 397$\,nm) and with 1/13 into the meta-stable $3^{2}$D$_{3/2}$-state (at $\lambda = 866$\,nm). In the presence of a magnetic field $\vec B$, due to $j$-mixing, there is a probability that the electron decays into the $3^{2}$D$_{5/2}$-state, with a branching ratio given by $4.2 \times 10^{-7}B^2 $~T$^{-2}$ \cite{Cric2010}. The population can be unshelved using the $4^{2}$P$_{3/2}$-state (at $\lambda = 854$\,nm). The absolute frequencies for these transitions in the absence of magnetic field are listed in Tab.~(\ref{absolute_frequencies}). \\

\begin{table}[b!]
\caption{Absolute frequencies of optical transitions of the $^{40}$Ca$^{+}$ ion. The saturation intensity $I_0$ is 433~$\mu $W/mm$^2$ for $^{2}$S$_{1/2}\,\rightarrow\,^{2}$P$_{1/2}$ transition and 3.4 and  3.3~$\mu $W/mm$^2$, for the transitions $^{2}$D$_{3/2}\,\rightarrow\,^{2}$P$_{1/2}$, and $^{2}$D$_{5/2}\,\rightarrow\,^{2}$P$_{3/2}$, respectively. $\Gamma $ is the decay rate (or linewidth) of the transition. ($^*$) This value is obtained from the frequency values of the $^{2}S_{1/2}\,\rightarrow\,^{2}P_{3/2}$ and $^{2}S_{1/2}\,\rightarrow\,^{2}D_{5/2}$ transitions quoted in Refs.~\cite{Wolf2009,Hua2015}.}
\vspace{0.2cm}
\centering
 \renewcommand{\arraystretch}{1.3}
\setlength{\tabcolsep}{1.mm}

	\begin{tabular}{ccc}
			\hline\hline
Atomic & Frequency & $\Gamma/2\pi$ \\
Transition & (MHz) & (MHz) \\
\hline \hline
			$^{2}$S$_{1/2}\,\rightarrow\,^{2}$P$_{1/2}$ & 755\,222\,766.2\,(1.7)  \cite{Wolf2008} & 21.57(8) \cite{Hett2015} \\
			$^{2}$D$_{3/2}\,\rightarrow\,^{2}$P$_{1/2}$ & 346\,000\,234.867\,(96)  \cite{Gebe2015}& 1.482(8) \cite{Hett2015} \\
			$^{2}$D$_{5/2}\,\rightarrow\,^{2}$P$_{3/2}$ & 350\,862\,882.830\,(91)  \cite{Koni2017}$^*$& 1.350(6) \cite{Gerr2008}  \\
	\hline \hline 
	\end{tabular} \label{absolute_frequencies}
\end{table}

\begin{figure*}[t!]
\hspace{0cm}
\centering\includegraphics[width=1.1\linewidth ]{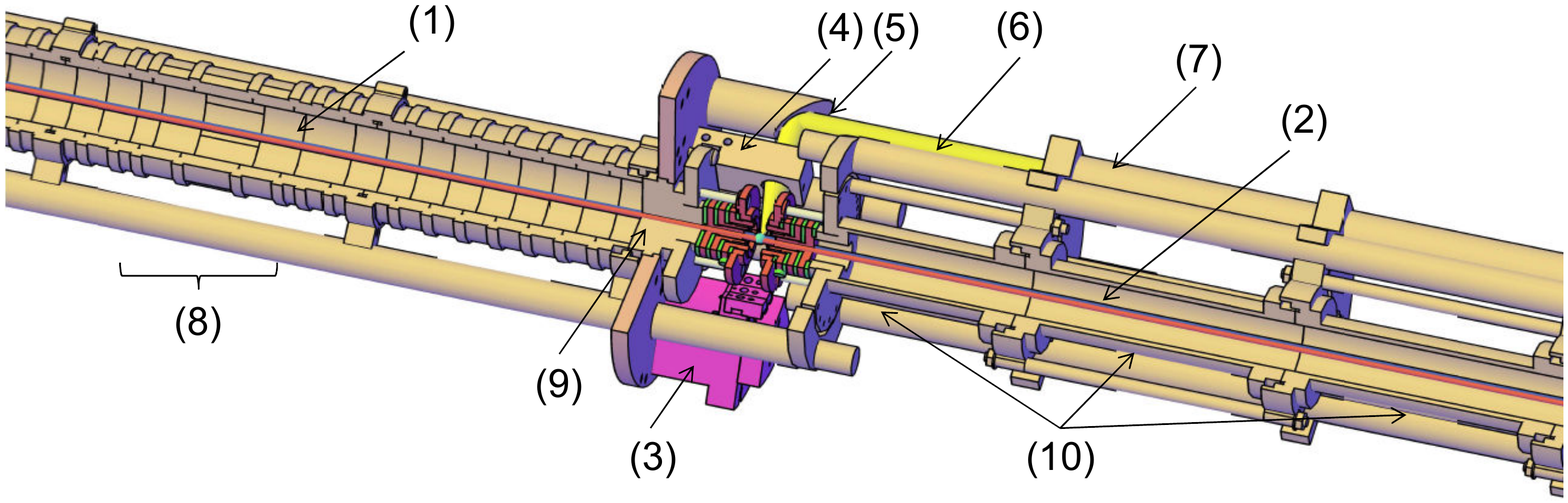}
\vspace{-7cm}
\caption{3-dimensional cut drawing showing part of the Penning-traps tower, together with the laser beams: 1) photoionization laser beams (crossing the MT from left to right), 2) laser beams for Doppler cooling (crossing the MT from right to left), 3) support structure for calcium ovens, 4)  support structure holding the first lens of the image collection system (f= 25~mm), 5) mirror of the image collection system, 6) surface depicting the size of the collimated fluorescence photons, 7) part of the structure housing the lenses inside vacuum, 8)  PT, 9) diaphragm, and 10) electrostatic lenses of the time-of-flight section for identification.}\label{Figure13}
\end{figure*}

\noindent In the presence of an external magnetic field $B$ the energy levels are no longer degenerated due to the linear Zeeman effect.  A detailed description and the resulting frequencies values to induce the different transitions for the observation of Doppler cooling in 7~Tesla are given in \ref{doppler_lasers}. \\

\begin{figure*}[t!]
\hspace{0cm}
\centering\includegraphics[width=1.1\linewidth ]{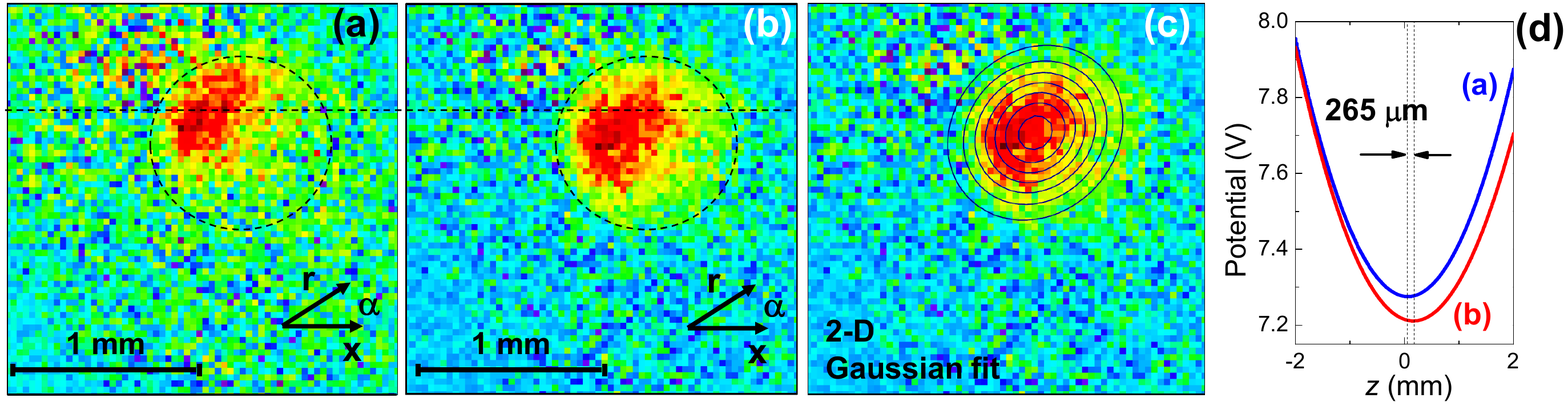}
\vspace{-8.2cm}
\caption{Images of the fluorescence photons from a $^{40}$Ca$^+$ ion cloud formed and stored in the 7~Tesla open-ring Penning trap. The detunings of the lasers were fixed to $\Delta _{\hbox{\scriptsize{397-nm}}}=-20$~MHz and $\Delta _{\hbox{\scriptsize{infrared}}}=-30$~MHz (regarding the values listed in Tab.~\ref{Tab:3} in \ref{doppler_lasers}). (a) and (b) differs in the potential applied to the endcap electrodes; in (a) the potential configuration is symmetric and in (b) non-symmetric. The two potential shapes along the axial direction are shown in (d) where the displacement of the trap center can be observed. The corresponding displacement between the images can be visualized following the dashed line and circumferences in (a) and (b). The acquisition time was 5~seconds, and each pixel in these images correspond to $16\times16$ pixels of the sensor. The plot labeled (c) shows a 2-dimenional Gaussian fit to the  image (b).} \label{Figure14}
\end{figure*}

\noindent The experimental laser arrangement for Doppler cooling comprises eight lasers with wavelengths 397~nm (x2), 866~nm (x4) and 854~nm (x2). The lasers are regulated by means of an ultra-accurate wavemeter with an absolute accuracy of 10~MHz. The lasers and control system as well as the outcomes from the regulation have been shown in previous works, i.e. in Refs. \cite{Corn2014} and \cite{Esco2014}. The sketch of the setup to generate the sidebands from the two carrier frequencies around the wavelength of 854~nm using an Electro Optical Modulator was presented in Ref.~\cite{Guti2016}. The laser beams are combined using standard optics elements, optical fibers, and they are all linearly polarized before entering the vacuum beamline. \\

\noindent Figure~\ref{Figure13} shows a three-dimensional cut drawing of part of the Penning-traps tower with the PT and the MT, indicating the axial laser beams for Doppler cooling. The figure shows the support structure made of MACOR, where three calcium ovens are placed. The laser beams for photoionization at $\lambda = 422$~nm and 375~nm, are also indicated in the figure. The resonant transition $^1$S$_0\rightarrow ^1$P$_{1}$ ($m_j=+1$) has been used (422~nm). The frequency for this transition is 98~GHz above the value used in the Paul trap experiments \cite{Domi2017}, due to the confinement  of the ion in the 7-Tesla magnetic field. 

\subsection{Fluorescence detection}

The fluorescence photons from the 4$^2$S$_{1/2}\rightarrow 4^2$P$_{1/2}$ transition in $^{40}$Ca$^+$ are registered by an Electron Multiplying Charge-Coupled Device (EMCCD) located approximately two meters downstream from the center of the open-ring trap. The sensor of the camera has $1024 \times 1024$ pixels, with a pixel size of 13~$\mu $m $\times $ 13~$\mu $m, thus having an active area of 13.3 $\times$ 13.3~mm$^2$. The optical system comprises a mirror, seven lenses (six of them inside vacuum, placed along 1.5 meters), and an objective with variable magnification. The positions of the first lens and the mirror are indicated in Fig.~\ref{Figure13}. The first lens has a focal distance of 25~mm and subtends a solid angle of about 1.6~$\%$. The set of all the lenses provides a fixed magnification factor, which combined with the objective might yield a further magnification by a factor of twelve. Figure~\ref{Figure14} shows images from the fluorescence photons emitted by a $^{40}$Ca$^+$ ion cloud interacting with the 12 laser beams. The $^{40}$Ca$^+$ ions are produced via photoionization inside the MT. In the first series of measurements the influence of each of the eight lasers (12 beams) has been observed, as well as the effect of the photoionization lasers.\\ 

\noindent Due to the orientation of the first lens and mirror (see Fig.~\ref{Figure13}), the fluorescence images of the cloud are rotated with respect to the axes of the camera, and therefore any analysis of the image in terms of their symmetry axes has to be performed doing a fit using 2-dimensional Gaussian functions (Fig.~\ref{Figure14}(b)), which can be written as

\begin{equation*}
f(x,y)= A\exp\left( {-\frac{[(x-x_0)\cos\alpha+(y-y_0)\sin\alpha]^2}{2\sigma_r^2}}\right )\times \
\end{equation*}
\begin{equation*}
\; \; \; \; \;  \; \; \; \; \; \; \; \; \; \; \; \;  \; \;  \exp\left( {-\frac{[(x-x_0)\sin\alpha+(y-y_0)\cos\alpha]^2}{2\sigma_z^2}}\right ),
\end{equation*}

\noindent where $A$ is the amplitude of the Gaussian (maximum photon counts), $\alpha$ is the angle formed between the radial plane and the horizontal direction in the EMCCD image (Fig.~\ref{Figure14}(a,b)). ($x_0,y_0$) is the center of the ion cloud obtained from the Gaussian fit, and $\sigma _z$ and $\sigma _r$ the standard deviation in the axial and radial direction, respectively. The maximum fluorescence signal has been obtained when the 397~nm lasers are detuned $-20$~MHz below resonance, and the detuning for all the infrared lasers is $\Delta =-30$~MHz (see Fig.~\ref{Figure14}). The fluorescence signal decreases for larger detuning of the 397-nm lasers, as observed when shifting the laser frequencies between $-100$ and $-300$~MHz. In the images shown in Fig.~\ref{Figure14}(a) and (b), the centers of the ion clouds obtained from the 2-dimensional Gaussian fits given in pixels are (34.6,18.6) and (37.1,21.8), respectively. This allows measuring the magnification factor from the displacement of the trap center (Fig.~\ref{Figure14}(d)), resulting in 3.2(5) when the variable magnification of the objective was set to the lowest value. The fluorescence distributions are different, resulting in $\sigma _z \sim 250$~$\mu$m and $\sigma _r \sim 320$~$\mu$m for the symmetric configuration (a), and $\sigma _z \sim 305$~$\mu$m and $\sigma _r \sim 355$~$\mu$m for the non-symmetric one (b).

\subsection{Doppler cooling}

Several simulations have been carried out to study the optimal settings to search for Doppler cooling. The simulations use the model from Ref.~\cite{Murb2016}, which in turn builds upon the laser cooling model presented in Ref.~\cite{Wese2007}, considering additionally the effects of the residual gas present in the trap. A brief overview will be presented here.\\

\noindent All the equations in the model consider only adimensional quantities. Thus, the energies are scaled by $ E_0 = \hbar \Gamma \sqrt{1+s_0} / 2 $ and the times by 
$ t_0 = 2 \left( 1+s_0 \right) / \left( \Gamma s_0 \right) $, where $ s_0 = I/I_0 $ is the saturation parameter. The relevant quantities are then

\begin{equation}
\epsilon = \frac{ E }{ E_0 }, \qquad \delta = \frac{ \hbar \Delta }{ E_0 }, \qquad r = \frac{ E_R }{ E_0 },  \qquad \mathrm{and} \qquad \tau = \frac{t}{t_0} ,
\end{equation}

\noindent where $ E $ is the ion's kinetic energy, $ \Delta $ is the detuning of the cooling laser and $ E_R = \left( \hbar k_z \right)^2 / 2 m  $ is the recoil energy of the ion when a photon is emitted. $ k_z $ is here the linear momentum of the cooling laser's photons.\\

\noindent Introducing $ Z = i / \sqrt{ 1 - \left( \delta + i \right)^2 / 4 \epsilon r } $, the rate equation for the ion's energy is \cite{Murb2016}

\begin{equation}
\frac{ d \epsilon }{ d \tau } = - 2 \delta_z \left( \epsilon - \epsilon_m \right) + \frac{4}{3} r \frac{1}{2 \sqrt{\epsilon r}} \mathrm{Im} \left( Z \right)
+ \frac{1}{2 \sqrt{\epsilon r}} \left[ \mathrm{Re} \left( Z \right) + \delta \mathrm{Im} \left( Z \right) \right]. \label{cooling_model}
\qquad 
\end{equation}

\begin{figure}[t!]
\hspace{0cm}
\centering\includegraphics[width=0.9\linewidth ]{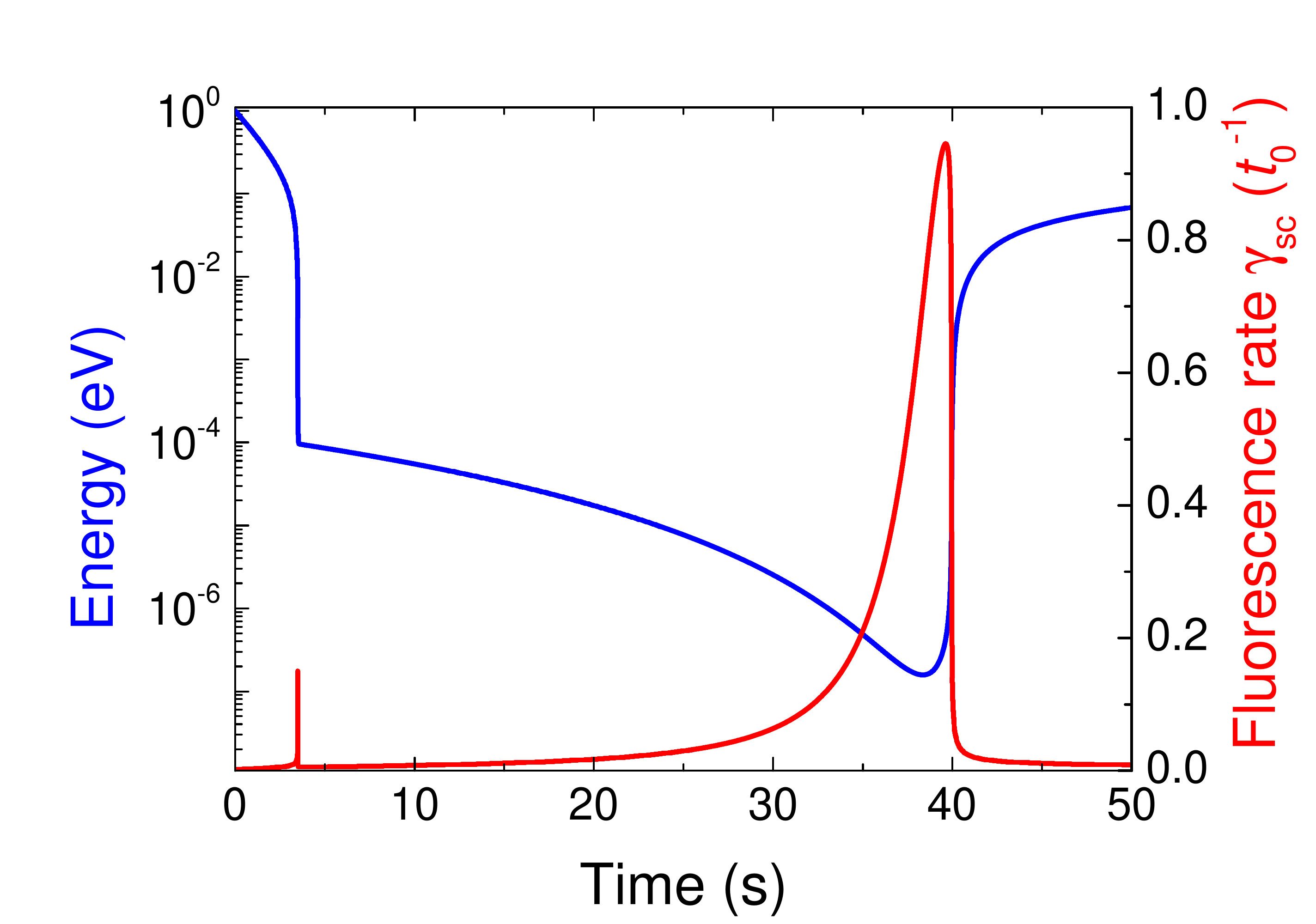}
\vspace{-0.4cm}
\caption{Doppler cooling of $^{40}$Ca$^+$ ions in helium buffer gas at 300~K. The background pressure was $\sim 10^{-7}$~mbar, and for this calculation only the $4^{2}$S$_{1/2}\rightarrow 4^{2}$P$_{1/2}$ has been considered ($\Gamma/2\pi =21.57(8)$~MHz). The saturation parameter $s_0$ was 0.4.}\label{Figure15}
\end{figure}

\noindent Here, the first term accounts for the effects of the residual gas, and the second and third terms describe the effects of the ion-laser interaction. $ \epsilon_m $ is the minimum energy attainable by buffer-gas cooling, and $ \delta_z $ was defined in Eq.~(\ref{Eq:GammeZ}).\\

\noindent The scaled fluorescence rate, in turn, is given by \cite{Murb2016}
\begin{equation}
\gamma_{sc} = \frac{ d N_{ph} }{ d \tau } = \frac{1}{2\sqrt{\epsilon r}} \mathrm{Im} \left( Z \right). \label{cooling_model_2}
\end{equation}

\noindent Figure~\ref{Figure15} shows the evolution of the ion's energy (left axis) and fluorescence rate (right axis) as a function of time, obtained by solving Eqs.~(\ref{cooling_model})~and~(\ref{cooling_model_2}). For this figure only the cooling transition in $^{40}$Ca$^+$ ($^{2}$S$_{1/2}\rightarrow ^{2}$P$_{1/2}$) has been considered, i.e., assuming a two-level system (like in the case of $^{24}$Mg$^+$ \cite{Murb2016}). However, this transition is not closed for $^{40}$Ca$^+$, where the population will have to be repumped from additional levels. The saturation parameters and detunings of these repumping drivings will affect the cooling rate in a non-trivial manner, and a more complete model will be needed to fully understand their effect \cite{Jana2015}. \\

\noindent Furthermore, we have observed another issue, which might prevent laser cooling. From the measured time-of-flight spectra, $^{40}$CaO$^+$  appears in the spectrum, besides $^{40}$Ca$^+$, and the ratio $^{40}$CaO$^+$/$^{40}$Ca$^+$ increases with the storage time in the MT. The expected background pressure in the MT was in the order of 10$^{-7}$~mbar for these measurements, and calcium oxide can only be formed in reactions, with residual atoms or molecules, since the $^{40}$Ca$^{+}$ ions are created resonantly. The vacuum is currently under improvement in order to minimize the presence of contaminants in the trap.\\

\section{Conclusions and Outlook}
At the University of Granada, we have built a full Penning-traps beamline setup to perform precision experiments using different ion species, in similar way as it is done at existing Penning-traps systems coupled to Radioactive Ion Beam facilities worldwide \cite{Kete2008,Bloc2005,Mukh2008,Ring2009,Webe2013,Kolh2004,Rodr2010}. However, the TRAPSENSOR facility has two outstanding features, i.e., 1) the novel geometry of the open-ring 7-Tesla Penning trap, which has been characterized in detail in this publication using destructive detection techniques, and 2) the lasers and fluorescence-image collection systems, built around the traps tower, to carry out laser spectroscopy. Moreover, the regime of 7-Tesla is also new to carry out laser-cooling experiments, as envisaged with this facility, to reach the first aim, namely to perform single-ion Penning trap mass spectrometry using a single laser-cooled ion as detector. Our ion choice to perform the first laser-cooling experiments is $^{40}$Ca$^+$, which needs twelve laser beams due to the large energy splitting arising from the Zeeman effect (first and second order). This does not occur in the other Penning-trap experiments of the same nature. In this publication, we have shown the first fluorescence measurement (photons with $\lambda =397$~nm), proving the interaction of the ions with the laser-cooling beams, which constitutes an important step towards Doppler-cooling of a single $^{40}$Ca$^+$ ion. Its subsequent application as a high-sensitive sensor will enable experiments in the framework of nuclear and fundamental physics. Such an implementation will be also interesting in the field of quantum technologies, for example for digital-analogue quantum simulations of spin models as Heisenberg \cite{Arra2016}, or spin-boson models as the Dicke model, in a variety of coupling regimes and inhomogeneities.\\

\section*{Acknowledgments}

This work was supported by the European Research Council (contract number 278648-TRAPSENSOR), from the Spanish MINECO/FEDER through [project number FPA2012-32076], [project number FPA2015-67694-P], [project number FIS2015-69983-P], [project number UNGR10-1E-501], [project number UNGR13-1E-1830], Ram\'on y Cajal Grant RYC-2012-11391, Juan de la Cierva grant IJCI-2015-26091, and "Programa de Garant\'ia Juvenil"; from the Spanish MECD through [PhD grant number FPU15-04679]; from Junta de Andaluc\'ia/FEDER IE-57131 and "Programa de Garant\'ia Juvenil"; from Basque Government [PhD grant number PRE-2015-1-0394] and [project number IT986-16], and from the University of Granada "Plan propio - Programa de Intensificaci\'on de la Investigaci\'on". We warmly thank Klaus Wendt for arranging the loan of the MALDI-TOF apparatus from the University of Mainz to the University of Granada.\\

\newpage

\appendix


\section{Frequencies of the cooling transitions for $^{40}$Ca$^+$ in 7~Tesla.} \label{doppler_lasers}
The Hamiltonian for a multielectronic atom in a magnetic field can be written as

\begin{equation}
H = H_0 + H' \,,
\end{equation}

\noindent where $ H_0 $ is the Hamiltonian in the absence of magnetic field (including fine structure) and

\begin{equation}
H' = \frac{ \mu_B B }{ \hbar } \left( L_z + g_S S_z \right) 
\end{equation}

\noindent is the Hamiltonian term due to the magnetic field. $\mu _B$ is the Bohr magneton \cite{CODATA}, $\hbar$ is the Planck`s constant divided by $2\pi$ \cite{CODATA}, and $g_S=2.002319304362(15)$ \cite{Odom2006}. For weak magnetic fields, i.e. when the corrections for the magnetic field are smaller than those of the fine structure, one can address $ H' $ as a perturbation. In the case of $^{40}$Ca$^+$, this holds true, e.g. the fine structure splitting of the $^2$P-state is 6.7~THz against $\Delta \nu$ ($^2$P$_{3/2}$)($_{\scriptsize{+3/2,-3/2}}$) $\approx 0.4$~THz. The first and second order correction to the energy of each unperturbed level $ \left| n^{ (0) } \right> $  are then given by

\begin{equation}
\Delta E^{ (1) } = \left< n^{ (0) } \right| H' \left| n^{ (0) } \right>
\end{equation}

\noindent and

\begin{equation}
\Delta E^{ (2) } = \sum_{ m \neq n } \frac{ \left| \left< m^{ (0) } \right| H' \left|  n^{ (0) } \right> \right| ^2 }{ E_n^{ (0) } - E_m^{ (0) } }
\,\,,
\end{equation}

\noindent respectively.\\

\noindent In order to compute these equations for each pair of atomic states, one must note that the eigenfunctions of the unperturbed Hamiltonian $ H_0 $ have well defined $ L $, $ S $, $ J $ and $ M_J$ quantum numbers, while the well defined numbers of the eigenfunctions of the perturbation $ H'$ are $ M_L $ and $ M_S $. The relationships between the states are given by the Clebsch-Gordan coefficients. \\

\begin{figure}[t!]
\centering
\includegraphics[width=1.0\textwidth]{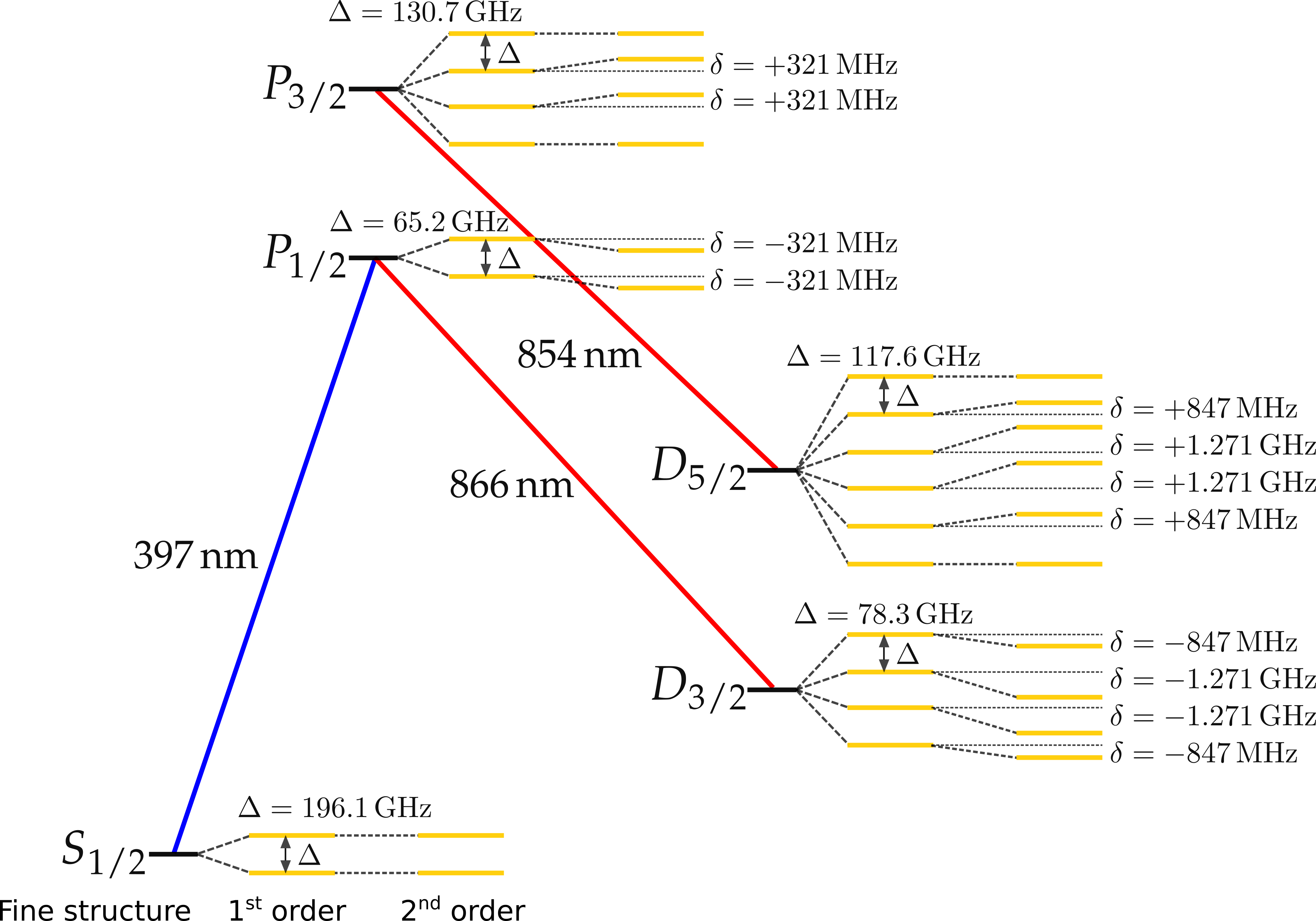}
\caption{Atomic level scheme of $^{40}$Ca$^+$ in a 7-T magnetic field. The figure shows the relevant transitions to perform Doppler cooling. First and second order effects of the magnetic field are included. Absolute frequencies are shown in Tab.~(\ref{absolute_frequencies}).}
\label{Figure12}
\end{figure}

\noindent In a strong magnetic field, the LS-coupling is no longer valid and one enters the regime of JJ-coupling. The discrepancy in the energy splitting of the P$_{1/2}$-state and the P$_{3/2}$-state (see e.g. Ref.~\cite{sobelman}) is below 1\,MHz at 7\,T when comparing the result with the energy shifts according to the linear and quadratic Zeeman effects. The calculated frequencies of the twelve laser beams are shown in Fig.~\ref{Figure12} and listed in Tab.~\ref{Tab:3}. \\

\begin{table}[t!]
\caption{Absolute frequencies of the transitions of the $^{40}$Ca$^{+}$ ion in a magnetic field $B=6.998747$\,T needed to perform Doppler cooling. The third column shows the standard deviations from the fixed frequency values during the regulation process. All laser beams are collimated with a diameter of $\approx 2$~mm. The laser powers were measured after they pass through the traps. For the 397-nm lasers the power was $1.6$ and $7.9$~mW for $\Delta m=+1$ and $\Delta m=-1$, respectively.  For the infrared lasers the values were 1-2~mW.($^*$) The lasers are set to a frequency 50~MHz above the resonant value.}

\vspace{0.2cm}
\centering
 \renewcommand{\arraystretch}{1.3}
\setlength{\tabcolsep}{3.2 mm}
\label{Tab:3}
	\begin{tabular}{ccc}
			\hline\hline
Atomic &  $\nu _{\hbox{\scriptsize{laser}}}$  & $\sigma _{\hbox{\scriptsize{laser}}}$\\
Transition &  (MHz) & (MHz)\\
\hline \hline
	$^{2}$S$_{1/2}\,\rightarrow\,^{2}$P$_{1/2}$    &   & \\  
	$m_{j}=-1/2\,\rightarrow\,m_{j}=+1/2$   & 755\,353\,131  & 1.17(1) \\ 
	$m_{j}=+1/2\,\rightarrow\,m_{j}=-1/2$   & 755\,091\,763 & 3.78(2) \\  \hline
		$^{2}$D$_{3/2}\,\rightarrow\,^{2}$P$_{1/2}$     & & \\ 
	$m_{j}=-3/2\,\rightarrow\,m_{j}=-1/2$   & 346\,085\,624  & 0.511(3) \\
	$m_{j}=-1/2\,\rightarrow\,m_{j}=+1/2$   &  346\,072\,955& 0.349(1) \\
	$m_{j}=+1/2\,\rightarrow\,m_{j}=-1/2$   &  345\,929\,407 & 0.297(1)\\ 
	$m_{j}=+3/2\,\rightarrow\,m_{j}=+1/2$   & 345\,915\,894 & 0.328(1) \\   \hline
	$^{2}$D$_{5/2}\,\rightarrow\,^{2}$P$_{3/2}$    &  &\\ 
	$m_{j}=-5/2\,\rightarrow\,m_{j}=-3/2$   & $350\,960\,839$  & \\ 
	$m_{j}=-3/2\,\rightarrow\,m_{j}=-1/2$  & 350\,973\,405$^*$ & 0.363(1) \\
	$m_{j}=-1/2\,\rightarrow\,m_{j}=+1/2$   & 350\,986\,075 &  \\ 
	$m_{j}=+1/2\,\rightarrow\,m_{j}=-1/2$  &  350\,737\,798 & \\
	$m_{j}=+3/2\,\rightarrow\,m_{j}=+1/2$   & 350\,751\,310$^*$ & 0.443(2) \\ 
	$m_{j}=+5/2\,\rightarrow\,m_{j}=+3/2$  &  $350\,764\,926$ & \\   \hline \hline
	\end{tabular}
\end{table}


\begin{thebibliography}{00}

\bibitem{Brow1986}
Brown L S and Gabrielse G 1986, Geonium theory: Physics of a single electron 
or ion in a Penning trap, {\it Rev. Mod. Phys.} {\bf 58} 233--311

\bibitem{Blau2013} 
Blaum K, Dilling J and N\"ortersh\"auser W 2013, Precision atomic physics techniques 
for nuclear physics with radioactive beams, {\it Phys. Scr.} {\bf T152} 014017

\bibitem{Repp2012} 
Repp J, B\"ohm C, Crespo L\'opez-Urrutia J R, D\"orr A, Eliseev S, George S, 
Goncharov M, Novikov Y N, Roux C, Sturm S, Ulmer U and Blaum K 2012, 
PENTATRAP: a novel cryogenic multi-Penning-trap experiment
for high-precision mass measurements on highly charged ions, 
{\it Appl. Phys. B: Lasers O.} {\bf 107} 983--996

\bibitem{Kete2008} Ketelaer J, Kr\"amer J, Beck D, Blaum K, Block M, 
Eberhardt K, Eitel G, Ferrer R, Geppert C, George S, Herfurth F, 
Ketter J, Nagy Sz, Neidherr D, Neugart R, N\"ortersh\"auser W, 
Repp J, Smorra C, Trautmann N and Weber C 2008, TRIGA-SPEC: A setup for 
mass spectrometry and laser spectroscopy at the research reactor TRIGA Mainz, 
{\it Nucl. Instrum. Methods A} {\bf 594} 162--177

\bibitem{Giac2017} 
Giacoppo F, Blaum K, Block M, Chhetri P, D\"ullmann Ch E, Droese C, Eliseev S, Filianin P, 
G\"otz S, Gusev Y, Herfurth F, Hessberger F P, Kaleja O, Khuyagbaatar J, Laatiaoui M, 
Lautenschl\"ager F, Lorenz C, Marx G, {Minaya Ramirez} E, Mistry A, Novikov Yu N, Plass W R, 
Raeder S, Rodr\'iguez D, Rudolph D, Sarmiento L G, Scheidenberger C, Schweikhard L, Thirolf P and 
Yakushev A 2017, Recent Upgrades of the SHIPTRAP Setup: On the Finish Line Towards Direct Mass 
Spectroscopy of Superheavy Elements, {\it Acta Physica Polonica B} {\bf 48} 423--429

\bibitem{Koni1995} 
K\"onig M, Bollen G, Kluge H-J, Otto T and Szerypo J 1995, Quadrupole excitation of stored ion motion 
at the true cyclotron frequency, {\it Int. J. Mass Spectrom. Ion Process.} {\bf 142} 95--116

\bibitem{Geor2007} 
George S, Baruah S, Blank B, Blaum K, Breitenfeldt M, Hager U, Herfurth F, 
Herlert A, Kellerbauer A, Kluge H-J, Kretzschmar M, Lunney D, Savreux R, 
Schwarz S, Schweikhard L and Yazidjian C 2007, Ramsey Method of Separated Oscillatory Fields
for High-Precision Penning Trap Mass Spectrometry, {\it Phys. Rev. Lett.} {\bf 98} 162501

\bibitem{Elis2013} 
Eliseev S, Blaum K, Block M, Droese C, Goncharov M, {Minaya-Ramirez} E,
Nesterenko D A, Novikov Yu N and Schweikhard L 2013, Phase-Imaging
Ion-Cyclotron-Resonance Measurements for Short-Lived Nuclides, {\it Phys. Rev. Lett.} {\bf 110} 082501

\bibitem{Bloc2010}
Block M, Ackermann D, Blaum K, Droese C, Dworschak M, Eliseev S,
Fleckenstein T, Haettner E, Herfurth F, Hessberger F P, Hofmann S,
Ketelaer J, Ketter J, Kluge H-J, Marx G, Mazzocco M, Novikov Yu N,
Plass W R, Popeko A, Rahaman S, {Rodr\'iguez} D, Scheidenberger C,
Schweikhard L, Thirolf P G, Vorobyev G K,¡ and Weber C 2010, Direct mass
measurements above uranium bridge the gap to the island of stability, {\it Nature} 
{\bf 463} 785--788

\bibitem{Elis2015} 
Eliseev S, Blaum K, Block M, Chenmarev S, Dorrer H, {D\"ullmann} Ch E,
Enss C, Filianin P E, Gastaldo L, Goncharov M, {K\"oster} U,
{Lautenschl\"ager} F, Novikov Yu N, Rischka A, {Sch\"ussler} R X,
Schweikhard L and {T\"urler} A 2015, Direct Measurement of the Mass Difference of  $^{163}$Ho and $^{163}$Dy Solves the $Q$-Value Puzzle for the Neutrino Mass Determination, {\it Phys. Rev. Lett.} {\bf 115} 062501

\bibitem{Mina2012} 
{Minaya Ramirez} E, Ackermann D, Blaum K, Block M, Droese C, {D\"ullmann} Ch E, Dworschak M, Eibach M, Eliseev S, Haettner E, Herfurth F, Hessberger F P, Hofmann S, Ketelaer J, Marx G, Mazzocco M, Nesterenko D, Novikov Yu N, Plass W, {Rodr\'iguez} D, Scheidenberger C, Schweikhard L, Thirolf P G, and Weber C 2012, Direct Mapping of Nuclear Shell Effects in the Heaviest Elements, {\it Science} {\bf 337} 1207--1210

\bibitem{Corn1989}
Cornell E A, Weisskoff R M, Boyce K R, {Flanagan Jr.} R W, Lafyatis G P and Pritchard D E 1989, Single-ion cyclotron resonance measurement of $ M$(CO$^{+}$)/$M$(N$^{+}_2$), {\it Phys. Rev. Lett.} {\bf 63} 1674--1677

\bibitem{Haef2003}
H\"affner H, Beier T, Djekic S, Hermanspahn N, Kluge H-J, Quint W, Stahl S, Verd\'u J, Valenzuela T and Werth G 2003, Double Penning trap technique for precise $g$-factor determinations in highly charged ions, {\it Eur. Phys. J. D} {\bf 22} 163-182

\bibitem{Vand2006}
{Van Dyck Jr.} R S, Pinegar D B, Liew S V and Zafonte S L 2006, The UW-PTMS: Systematic studies, measurement progress, and future improvements, {\it Int. J. Mass Spectrom.} {\bf 251} 231--242

\bibitem{Stur2014}
Sturm S, {K\"ohler} F, Zatorski J, Wagner A, Harman Z, Werth G, Quint W, Keitel C H and Blaum K 2014, High-precision measurement of the atomic mass of the electron, {\it Nature} {\bf 506} 467--470

\bibitem{Ulme2015} 
Ulmer S, Smorra C, Mooser A, Franke K,  Nagahama H, Schneider G, Higuchi T, Van~Gorp S, Blaum K, Matsuda Y, Quint W, Walz J and Yamazaki Y 2015, High-precision comparison of the antiproton-to-proton charge-to-mass ratio, {\it Nature} {\bf 524} 196--199

\bibitem{Heis2017}
Heisse F, K\"ohler-Langes F, Rau S, Hou J, Junck S, Kracke A, Mooser A, Quint W, Ulmer S, Werth G, Blaum K and Sturm S 2017, High-Precision Measurement of the Proton’s Atomic Mass, {\it Phys. Rev. Lett.} {\bf 119}  033001

\bibitem{Mina2013}
{Minaya Ramirez} E, Ackermann D, Blaum K, Block M, Droese C, {D\"ullmann} Ch E, Eibach M, Eliseev S, Haettner E, Herfurth F, Hessberger F P, Hofmann S, Marx G, Nesterenko D, Novikov Yu N, Plass W, {Rodr\'iguez} D, Scheidenberger C, Schweikhard L, Thirolf P G and Weber C 2013, Recent developments for high-precision mass measurements of the heaviest elements at SHIPTRAP, {\it Nucl. Instrum. Methods B} {\bf 317} 501--505

\bibitem{Rodr2012} 
{Rodr\'iguez} D 2012, A quantum sensor for high-performance mass spectrometry,
{\it Appl. Phys. B: Lasers O.} {\bf 107} 1031--1042

\bibitem{Hein1990} 
Heinzen D J and Wineland D J 1990, Quantum-limited cooling and detection of radio-frequency oscillations by laser-cooled ions, {\it Phys. Rev. A} {\bf 42(5)} 2977--2994

\bibitem{Corn2015}
Cornejo J M, Colombano M, {Dom\'enech} J, Block M, Delahaye P and {Rodr\'iguez} D 2015, Extending the applicability of an open-ring trap to perform experiments with a single laser-cooled ion, {\it Rev. Sci. Instrum.} {\bf 86} 103104

\bibitem{Domi2017}
Dom\'inguez F, Arrazola I, Dom\'enech J, Pedernales J S, Lamata L, Solano E and Rodr\'iguez D 2017, A Single-Ion Reservoir as a High-Sensitive Sensor of Electric Signals, {\it Sci. Rep.} {\bf 7} 8336

\bibitem{Domi2017b}
Dom\'inguez F, Guti\'errez M J, Arrazola I, Berrocal J, Cornejo J M, Del~Pozo J J, Rica R A, Schmidt S, Solano E and Rodr\'iguez D 2018, Motional studies of one and two laser-cooled trapped ions for electric-field sensing applications, {\it J. Mod. Opt.} {\bf 65} 613--621

\bibitem{Drew2004}
Drewsen M, Mortensen A, Martinussen R, Staanum P and S\o rensen J L 2004, Nondestructive Identification of Cold and Extremely Localized Single Molecular Ions, {\it Phys. Rev. Lett.} {\bf 93} 243201

\bibitem{Laat2016}
Laatiaoui M, Lauth W, Backe H, Block M, Ackermann D, Cheal B, Chhetri P, D\"ullmann Ch E, Van~Duppen P, Even J, Ferrer R, Giacoppo F, G\"otz S, Hessberger F P, Huyse M, Kaleja O, Khuyagbaatar J, Kunz P, Lautenschl\"ager F, Mistry A K, Raeder S, Minaya Ramirez E, Walther T, Wraith C and Yakushev A 2016, Atom-at-a-time laser resonance ionization spectroscopy of nobelium, {\it Nature} {\bf 538} 495--498

\bibitem{Schm2005}
Schmidt P O, Rosenband T, Langer C, Itano W M, Bergquist J C and Wineland D J 2015, 
Spectroscopy Using Quantum Logic, {\it Science} {\bf 309} 749--752 

\bibitem{Ospe2014}
Niemann M, Paschke A G, Dubielzig T, Ulmer S and Ospelkaus C 2014, CPT Test with
(anti)proton Magnetic Moments Based on Quantum Logic Cooling and Readout.
In CPT and Lorentz Symmetry-Proceedings of the Sixth Meeting. Vol. 1, Bloomington,
Indiana, USA 41--44

\bibitem{Bloc2005} 
Block M, Ackermann D, Beck D, Blaum K, Breitenfeldt M, Chauduri A, Doemer A, Eliseev S, 
Habs D, Heinz S, Herfurth F, Heßberger F P, Hofmann S, Geissel H, Kluge H-J, Kolhinen V, 
Marx G,  Neumayr J B, Mukherjee M, Petrick M, Plass W R, Quint W, Rahaman S, Rauth C, 
Rodr\'iguez D, Scheidenberger C, Schweikhard L, Suhonen M, Thirolf P G, Wang Z and Weber C 2005, 
The ion-trap facility SHIPTRAP, {\it Eur. Phys. J. A} {\bf 25} 49--50

\bibitem{Mukh2008}
Mukherjee M, Beck D, Blaum K, Bollen G, Dilling J, George S, Herfurth F, Herlert A, 
Kellerbauer A, Kluge H-J, Schwarz S, Schweikhard L and Yazidjian C 2008, ISOLTRAP: An on-line 
Penning trap for mass spectrometry on short-lived nuclides, {\it Eur. Phys. J. A} {\bf 35} 1--29

\bibitem{Ring2009} 
Ringle R, Bollen G, Prinke A, Savory J, Schury P, Schwarz S and Sun T 2009, The LEBIT 9.4 T Penning 
trap mass spectrometer, {\it Nucl. Instrum. Methods A} {\bf 604} 536--547

\bibitem{Corn2013}
Cornejo J M, Lorenzo A, Renisch D, Block M, {D\"ullmann} Ch E and {Rodr\'iguez} D 2013, Status of the project TRAPSENSOR: Performance of the laser-desorption ion source, {\it Nucl. Instrum. Methods B} {\bf 317} 522--527

\bibitem{Webe2013} Weber C, M\"uller P and Thirolf P G 2013,  Developments in Penning trap (mass) spectrometry at MLLTRAP:
Towards in-trap decay spectroscopy, {\it Int. J. Mass Spectrom.} {\bf 349--350} 270--276

\bibitem{Kolh2004} Kolhinen V S, Kopecky S, Eronen T, Hager U, Hakala J, Huikari J, Jokinen A, Nieminen A, Rinta-Antila S, Szerypo J, \"Ayst\"o J, JYFLTRAP: a cylindrical Penning trap for isobaric beam purification at IGISOL, 
{\it Nucl. Instrum. Methods A} {\bf 528} 776--787

\bibitem{Rodr2010} Rodr\'iguez D, Blaum K, N\"ortersh\"auser W, Ahammed M, Algora A, Audi G, \"Ayst\"o J $et$ $al$ 2010, MATS and LaSpec: High-precision experiments using ion traps and lasers at FAIR, {\it Eur. Phys. J. Special Topics} {\bf 183} 1--123

\bibitem{Sava1991}
Savard G, Becker St, Bollen G, Kluge H-J, Moore R B, Otto Th, Schweikhard L, Stolzenberg H and Wiess U 1991, A new cooling technique for heavy ions in a Penning trap, {\it Phys. Lett. A} {\bf 258} 247--252

\bibitem{Gabr1996}
Hall D S and Gabrielse G 1996, Electron Cooling of Protons in a Nested Penning Trap, {\it Phys. Rev. Lett.} {\bf 77} 1962--1965

\bibitem{Corn2016a} 
Cornejo J M, Guti\'errez M J, Ruiz E, Bautista-Salvador A, Ospelkaus C, Stahl S and Rodr\'iguez D 2016, 
An optimized geometry for a micro Penning-trap mass spectrometer based on interconnected ions,
{\it Int. J. Mass Spectrom.} {\bf 410} 22--30

\bibitem{Rica2018}
Rica R A, Dom\'inguez F, Guti\'errez M J, Ba\~nuelos J, Del Pozo J J and Rodr\'iguez D 2018, A double Paul trap system for the electronic coupling of ions, accepted for publication in the {\it Eur. Phys. J. Special Topics}.

\bibitem{Buss2006}
Bussmann M, Schramm U, Habs D, Kolhinen V S and Szerypo J 2006, Stopping highly charged ions in a laser-cooled one component plasma of $^{24}$Mg$^+$ ions, {\it Int. J. Mass Spectrom.} {\bf 251} 17--189 

\bibitem{Schm2015}
Schm\"oger L, Versolato O, Schwarz M, Kohnen M, Windberger A, Piest B, Feuchtenbeiner S, Pedregosa-Gutierrez J, Leopold T, Micke P, Hansen A K, Baumann T M, Drewsen M, Ullrich J, Schmidt P O and Crespo~López-Urrutia 2015 J R, Coulomb crystallization of highly charged ions, {\it Science} {\bf 347} 1233-1236

\bibitem{Schm2017} 
Schmidt S, Murb\"ock T, Andelkovic Z, Birkl G, K\"onig K, N\"ortersh\"auser W, Thompson R C and Vogel M 2017, Sympathetic cooling in two-species ion crystals in a Penning trap, {\it J. Mod. Opt.} {\bf 65} 538--548

\bibitem{Corn2016b}
Cornejo J M and {Rodr\'iguez} D 2016, A preparation penning trap for the TRAPSENSOR project with prospects for MATS at FAIR, {\it Nucl. Instrum. Methods B} {\bf 376} 288--291

\bibitem{Corn2016c}
Cornejo J M 2016, The Preparation Penning Trap and Recent Developments on High-Performance Ion Detection for the Project TRAPSENSOR, Ph.D. thesis, Universidad de Granada

\bibitem{Elli1984}
Ellis H W, Thackston M G, McDaniel E W, Mason E A 1984, Transport properties of gaseous ions 
over a wide energy range. Part III, {\it At. Data Nucl. Data Tables} {\bf 31} 113--151

\bibitem{Rodr2006}
Rodr\'iguez D, M\'ery A, Ban G, Bregeault J, Darius G, Durand D, Fl\'echard X, Herbane M, Labalme M, Li\'enard E, Mauger F, Merrer Y, Naviliat-Cuncic O, Thomas J C, Vandamme C 2006, The LPCTrap facility: a novel transparent Paul trap for high-precision experiments, {\it Nucl. Instrum. Methods A} {\bf 565} 476--489

\bibitem{Flec2008}
Fl\'echard X, Li\'enard E, M\'ery A, Rodr\'iguez D, Ban G, Durand D, Duval F, Herbane M, Labalme M, Mauger F, Naviliat-Cuncic O and Thomas J C 2008, Paul trapping of radioactive $^6$He$^+$ ions and direct observation of their beta decay, {\it Phys. Rev. Lett.} {\bf 101} 212504

\bibitem{Flec2011}
Fl\'echard X, Velten Ph, Li\'enard E, M\'ery A, Rodr\'iguez D, Ban G, Durand D, Mauger F, Naviliat-Cuncic O and Thomas J C 2011, Measurement of the $\beta$-$\nu$ correlation coefficient $a_{\scriptsize{\beta \nu}}$ in the $\beta$ decay of trapped $^6$He$^+$ ions, {\it J. Phys. G: Nucl. Part. Phys.} {\bf 38} 055101

\bibitem{SIMION}
\url{http://simion.com/}

\bibitem{Cric2010} 
Crick D R, Donnellan S, Segal D M and Thompson R C 2010, Magnetically induced electron shelving
 in a trapped Ca$^+$ ion, {\it Phys. Rev. A} {\bf 81} 052503

\bibitem{Wolf2008}
Wolf A L, van den Berg S A, Gohle C, Salumbides E J, Ubachs W and Eikema K S E 2008, 
Frequency metrology on the 4s $^2$S$_{1/2}$–4p $^2$P$_{1/2}$ transition in $^{40}$Ca$^+$ for a comparison with quasar data, {\it Phys. Rev. A} {\bf 78} 032511

\bibitem{Hett2015} Hettrich M, Ruster T, Kaufmann H, Roos C F, Schmiegelow C T, Schmidt-Kaler F and Poschinger U G 2015, 
Measurement of Dipole Matrix Elements with a Single Trapped Ion, {\it Phys. Rev. Lett.} {\bf 115} 143003

\bibitem{Gebe2015} Gebert F, Wan Y, Wolf F, Angstmann C N, Berengut J C and Schmidt P O 2015,  Precision Isotope Shift Measurements in Calcium Ions Using Quantum Logic Detection Schemes, {\it Phys. Rev. Lett.} {\bf 115} 053003

\bibitem{Koni2017}
K\"onig K, Geppert Ch,  Kr\"amer J, Maass B, Otten E W, Ratajczyk T and N\"ortersh\"auser W 2017, First high-voltage measurements using Ca$^+$ ions at the ALIVE experiment, {\it Hyperfine Interact.} {\bf 238} 24

\bibitem{Wolf2009} Wolf A L, van~den~Berg S A, Ubachs W and  Eikema K S E 2009, Direct Frequency Comb Spectroscopy of Trapped Ions, {\it Phys. Rev. Lett.} {\bf 102} 223901

\bibitem{Hua2015} 
 Guan H, Huang Y, Liu P-L, Bian W, Shao H and Gao K L 2015,  Precision spectroscopy with a single $^{40}$Ca$^+$ ion in a Paul trap, {\it Chinese Phys. B} {\bf 24} 054213

\bibitem{Gerr2008} Gerritsma R, Kirchmair G, Z\"ahringer F, Benhelm J, Blatt R and Roos CF 2008, Precision measurement of the branching fractions of the 4p $^2$P$_{3/2}$ decay of Ca II, {\it Eur. Phys. J. D} {\bf 50} 13--19

\bibitem{CODATA}
\url{https://physics.nist.gov/cuu/Constants/}

\bibitem{Odom2006} Odom B, Hanneke D, D\ {'}Urso B and Gabrielse G 2006,
New Measurement of the Electron Magnetic Moment Using a One-Electron Quantum Cyclotron,  
{\it Phys. Rev. Lett.} {\bf 97} 030801

\bibitem{sobelman}
 Sobelman I I 1979, Atomic Spectra and Radiative Transitions, Berlin, Springer-Verlag

\bibitem{Corn2014} Cornejo J M, Escobedo P and Rodr\'iguez D (2014), 
Status of the project TRAPSENSOR, {\it Hyperfine Interact.} {\bf 227} 223--237 

\bibitem{Esco2014}
Escobedo P 2014, Desarrollo de un sistema de control para l\'aseres de diodo utilizando moduladores ac\'ustico-\'opticos, Master Thesis, Universidad de Granada

\bibitem{Guti2016}
Guti\'errez M J 2016, Estudio de la fluorescencia de un ion de $^{40}$Ca$^+$ en una trampa magn\'etica de 7 T, Master Thesis, Universidad de Granada

\bibitem{Murb2016} Murb\"ock T, Schmidt S, Birkl G, N\"orterh\"auser W, Thompson R C and Vogel M 2016,
Rapid crystallization of externally produced ions in a Penning trap, {\it Phys. Rev. A} {\bf 94} 043410

\bibitem{Wese2007} Wesenberg J H, Epstein R J, Leibfried D, Blakestad R B, Britton J, Home J P, Itano W M, Jost J D,
Knill E, Langer C, Ozeri R, Seidelin S, Wineland D J 2007, Fluorescence during Doppler cooling of a single trapped atom, {\it Phys. Rev. A} {\bf 76} 053416

\bibitem{Jana2015}
Janacek H A 2015, Optical Bloch equations for simulating trapped-ion qubits, Ph.D. thesis, University of Oxford 

\bibitem{Arra2016} 
Arrazola I, Pedernales J S, Lamata L and Solano E 2016, Digital-Analog Quantum Simulation of Spin Models in Trapped Ions, 
{\it Sci. Rep.} {\bf 6} 30534

\end{thebibliography}
\end{document}